\documentclass[aps,prd,twocolumn,floatfix,nofootinbib,superscriptaddress,tightenlines]{revtex4}

\usepackage[dvipsnames]{xcolor}
\usepackage{amsmath}
\usepackage{dcolumn}
\usepackage{lipsum}
\usepackage{amssymb}
\usepackage{url}
\usepackage{epsfig}
\usepackage{graphicx}
\usepackage{amsmath}
\usepackage{bm}
\usepackage{setspace}
\usepackage{lscape}
\usepackage{amsthm}
\usepackage{bbold}
\usepackage{dcolumn}
\usepackage{epsfig}
\usepackage{graphics}
\usepackage{graphicx}
\usepackage{longtable}

\usepackage{bm}
\usepackage{xspace}
\usepackage{cancel}
\usepackage{float}
\usepackage{multirow}
\definecolor{darkgreen}{rgb}{0,0.5,0}
\definecolor{purple}{rgb}{0.5,0,0.5}
\definecolor{nblue}{rgb}{0.0,0.0,0.50}
\definecolor{scarlet}{rgb}{1.0,0.2,0}
\definecolor{darkmagenta}{rgb}{0.55, 0.0, 0.55}
\definecolor{darkolivegreen}{rgb}{0.33, 0.42, 0.18}
\definecolor{darkcandyapplered}{rgb}{0.64, 0.0, 0.0}

\usepackage[colorlinks=true, pdfstartview=FitV, linkcolor=purple, citecolor= purple, urlcolor=blue]{hyperref}

\newcommand{\be}{\begin{equation}}
\newcommand{\tu}{\textcolor{red}{u}}

\newcommand{\f}{\textcolor{blue}{f}}
\newcommand{\fu}{\textcolor{blue}{\bar{f_2}}}
\newcommand{\fd}{\textcolor{blue}{f_1}}
\newcommand{\fdu}{\textcolor{blue}{f_2}}

\newcommand{\Meps}{\textcolor{blue}{{PS}}}
\newcommand{\Ms}{\textcolor{blue}{{S}}}

\newcommand{\td}{\textcolor{darkcandyapplered}{d}}
\newcommand{\tb}{\textcolor{blue}{b}}
\newcommand{\tc}{\textcolor{darkmagenta}{c}}
\newcommand{\ts}{\textcolor{darkgreen}{s}}
\newcommand{\ee}{\end{equation}}
\newcommand{\bea}{\begin{eqnarray}}
\newcommand{\eea}{\end{eqnarray}}
\newcommand{\beas}{\begin{eqnarray*}}
\newcommand{\eeas}{\end{eqnarray*}}
\newcommand{\nn}{\nonumber}

\newcommand{\MeV}{\text{MeV}} 
\newcommand{\GeV}{\text{GeV}} 

\newcommand{\tab}[1]{Table~\ref{#1}}

\begin{document}
\title{Electromagnetic Form Factors and Charge Radii of Pseudoscalar and Scalar Mesons: A Comprehensive Contact Interaction Analysis}
\author{R. J. Hern\'andez-Pinto}
\email{roger@uas.edu.mx}
\affiliation{Facultad de Ciencias F\'isico-Matem\'aticas, Universidad Aut\'onoma de Sinaloa, Ciudad Universitaria, Culiac\'an, Sinaloa 80000,
M\'exico}

\author{L. X. Guti\'errez-Guerrero}
\email{lxgutierrez@conacyt.mx}
\affiliation{CONACyT-Mesoamerican Centre for Theoretical Physics,
Universidad Aut\'onoma de Chiapas, Carretera Zapata Km.~4, Real
del Bosque (Ter\'an), Tuxtla Guti\'errez, Chiapas 29040, M\'exico}

\author{A. Bashir}
\email{adnan.bashir@umich.mx;
 abashir@jlab.org}

\affiliation{Instituto de F\'isica y Matem\'aticas, Universidad
Michoacana de San Nicol\'as de Hidalgo, Edificio C-3, Ciudad
Universitaria, Morelia, Michoac\'an 58040, M\'exico}
\affiliation{Thomas Jefferson National Accelerator Facility, Newport News, Virginia 23606, USA}

\author{M. A. Bedolla}
\email{marco.bedolla@unach.mx}
\affiliation{Facultad de Ciencias en Física y Matemáticas, Universidad Aut\'onoma de Chiapas, 
Carretera Emiliano Zapata Km.~8, Rancho San Francisco, Ciudad Universitaria Ter\'an, Tuxtla Guti\'errez, Chiapas 29040, M\'exico}

\author{I. M. Higuera-Angulo}
\email{melany.higuera@umich.mx}
\affiliation{Instituto de F\'isica y Matem\'aticas, Universidad
Michoacana de San Nicol\'as de Hidalgo, Edificio C-3, Ciudad
Universitaria, Morelia, Michoac\'an 58040, M\'exico}

\begin{abstract}

We carry out a comprehensive survey of electromagnetic form factors of all light, heavy and heavy-light ground-state pseudoscalar and scalar mesons. Our analysis is based upon a Schwinger-Dyson equations treatment of a
vector $\times$ vector contact interaction. It incorporates confinement and ensures axial vector and vector Ward-Takahashi identities are satisfied along with the corresponding corollaries such as the Goldberger-Treiman relations. 
The algebraic simplicity of the model allows us to compute the
form factors at arbitrarily large virtualities of the probing photon momentum squared with relative ease. Wherever possible and insightful, we compare our results for the electromagnetic form factors and the charge radii with those obtained earlier through Schwinger-Dyson equations, lattice and with experimental observations available. We also comment on the scope and shortcomings of the model. 

\end{abstract}


\maketitle

\section{Introduction}


A major challenge in strong interaction physics is the description of hadrons from first principles, i.e., by commencing from the Lagrangian dynamics of elementary degrees of freedom of quantum chromodynamics (QCD), namely, quarks and gluons. The arduous task is then to describe hadron properties by sewing together the Green functions of dressed quarks through relativistic bound state equations. In close analogy with the hydrogen atom of electrodynamics, the simplest bound states of QCD are the two-particle systems (mesons) composed of a quark and an antiquark ($q\bar{q}'$). Relativistic description of such states through the Bethe-Salpeter equation (BSE) was first formulated in Ref.~\cite{Salpeter:1951sz}. 
Solutions of this equation presuppose the knowledge of the dressed quark propagator and the $q\bar{q}'$ scattering kernel. The quark propagator is obtained by solving the gap equation while the $q\bar{q}'$ scattering kernel is constructed by ensuring the axial vector Ward-Takahashi identity is satisfied. 

Several experimental facilities around the globe study electromagnetic properties of mesons for a gradually increasing interval of momentum  squared ($Q^2$) transferred to the target by the incident probing photon. It enhances the possibility of observing a gradual transition from non-perturbative QCD effects to its perturbative domain, finally settling onto its asymptotic predictions estimated decades ago, all that in one single experiment. Resulting elastic or electromagnetic form factors (EFFs) of mesons thus provide us with an ideal platform to study numerous uncanny facets of QCD, unfolding the complex structure of these bound states at varying resolutions scales. 

EFFs of pseudoscalar (PS) mesons, pion and kaon in particular but, have been studied extensively, for example, within the functional approach via Schwinger-Dyson
equations (SDEs)~\cite{Maris:2000sk,Chang:2013nia,Bhagwat:2006pu}, lattice QCD~\cite{Shultz:2015pfa,Alexandrou:2021ztx,Gao:2021xsm,Davies:2018zav}, contact interaction (CI)~\cite{Gutierrez-Guerrero:2010waf,Wang:2022mrh,Xing:2022jtt}, other models and formalisms, asymptotic QCD~\cite{Lepage:1979zb} and, of course,
experimentally~\cite{NA7:1986vav,JeffersonLabFpi-2:2006ysh,JeffersonLabFpi:2000nlc}.  $\pi^+$, $K^+$, $K^0$, $D^+$ and $D_0$ have also been studied in a hybrid model that combines the generalized
Bertlmann-Martin inequalities with smearing corrections due to relativistic effects~\cite{Lombard:2000kw}. Light and heavy PS mesons in the  light-front framework  have been reported in~\cite{Hwang:2001th} while with a QCD potential model there are results for $D^+$, $D^0$, $D_s^+$, $B^+$, $B^0$, $B^0_{\ts}$~\cite{Das:2016rio}.

PS mesons have additional and important relevance as they contribute to 
the hadronic light-by-light (HLbL) piece of
the muon anomalous magnetic moment (AMM), most dominantly through the single exchange of the light mesons such as $\pi, \eta$ and $\eta'$. Furthermore, there are also loops with charged pions ($\pi^{\pm}$) and kaons ($K^{\pm}$). These contributions have been computed with desirable accuracy within the SDE formalism~\cite{Goecke:2010if,Eichmann:2019tjk,Raya:2019dnh,Eichmann:2019bqf,Miramontes:2021exi}.
On the other hand, scalar (S) mesons have been less studied for technical hindrances and due to the fact their composition is still debatable. However, similarly to the PS mesons, they contribute to the AMM of the muon, see the review article~\cite{Aoyama:2020ynm} and references therein. 

 Additional overwhelming interest in studying mesons arises from the fact that their BSE analysis provides an important first step towards studying baryons in a quark-diquark picture.  It is firmly established that the non-pointlike diquark correlations play an important role in baryons~\cite{Barabanov:2020jvn}. As a clear illustration, it has been demonstrated that the quark-diquark  picture of a nucleon produces its mass within 5\% of what the Faddeev equation of a three quark system~\cite{Eichmann:2011vu} yields. With this realisation, 
 it is useful to know that the BSE for diquarks is exactly the same as that for corresponding mesons up to a color and charge factor. The chiral partners
form a set of particles which transform into each other under chiral transformation, like $(\sigma, \pi)$ and $(\rho,a_{1})$. Correspondingly, there are diquark partners 
($0^-,0^+$) and $(1^+,1^-)$.
The Bethe-Salpeter amplitudes (BSAs) as well as the EFFs for $\sigma, \pi, \rho$ and $a_1$ yield the corresponding description of diquarks $0^-,0^+,1^+$ and $1^-$.
The quark-diquark picture has been successfully used to calculate EFFs and transition form factors (TFF) of baryons~\cite{Cloet:2008re,Wilson:2011aa,Segovia:2013uga,Segovia:2014aza,Raya:2018ioy,Raya:2021pyr,Segovia:2015hra}. For comprehensive reviews in this connection, one can consult Refs.~\cite{Aznauryan:2012ba,Bashir:2012fs}.

We have already mentioned CI in the preceding discussion.  It is a symmetry preserving  vector × vector interaction
based on a momentum-independent gluon propagator. It results in four quarks interacting at a point.
It was first proposed in~\cite{Gutierrez-Guerrero:2010waf}
to calculate pion EFF. Subsequently, CI has extensively been employed to study EFFs and TFFs of mesons
in Refs.~\cite{Roberts:2010rn,Roberts:2011wy,Chen:2012txa,Raya:2017ggu,Wang:2022mrh} and of baryons in Refs.~\cite{Wilson:2011aa,Segovia:2013uga,Segovia:2014aza,Raya:2018ioy,Raya:2021pyr}.
It is a well-known realization that the EFFs obtained from the CI are harder than the ones obtained from full QCD predictions. However, the simplicity of the model allows us to perform algebraic calculations. Moreover, the results obtained provide a benchmark to compare and contrast with refined QCD-based SDE results in order to understand the correct pattern of dynamical chiral symmetry breaking (DCSB) and the large $Q^2$ evolution of the EFFs which stems from asymptotic QCD where $Q^2$ is much larger than any other mass scale relevant to the problem. In this work, we compute EFFs using this momentum-independent interaction, regularized in a symmetry-preserving manner for a large number of PS and S mesons composed of light quarks, heavy quarks, and the heavy-light combinations. We must emphasize that the scalars like $\sigma$ have a complicated internal structure, possibly including a large component of pion correlations. The $\sigma$ in our article refers to a quark-antiquark state alone, parity partner of the pion and approximately twice as heavy as $\sigma(600)$,~\cite{Pelaez:2006nj}.

The article is organized as follows: in Sec.~\ref{BSE} we collect the basic ingredients required to
carry out the analysis in the CI model: the dressed quark masses obtained through the gap equation and the general expression for the BSAs for PS and S mesons.
We discuss the generalities of the EFFs for PS and S mesons in Sec.~\ref{sec:eff}, i.e., the quark-photon vertex and the triangle diagram which are the two building blocks to calculate all the meson EFFs in our formalism. Sec.~\ref{PS-S-FF} is dedicated to computing EFFs of the ground state PS mesons. It allows us to evaluate their charge
radii in the limit $Q^2 \approx  0$, and simultaneously understand the asymptotic behaviour of the meson EFFs at large $Q^2$, i.e., $Q^2 \rightarrow \infty$. In Sec.~\ref{S-FF}, we repeat our study for the S mesons. 
A brief summary and perspectives for future work are presented in
Sec.~\ref{Conclusions}. 

\section{The Ingredients} \label{BSE}
Calculation of the meson EFFs presupposes the
knowledge of the dynamically generated dressed valence quark masses, BSAs of the mesons as
well as the quark-photon interaction vertex at different probing momenta of the incident photon. In this section, we provide a brief but self contained introduction to the CI, its essential ingredients and characteristics, namely, the gluon propagator, the quark-gluon vertex and the set of parameters employed which, collectively, define the CI. This discussion is followed by the solution of the gap equation to obtain dynamically generated dressed quark masses. We then provide the general expressions of the BSAs for PS and S mesons. The corresponding BSE is set up consistently with the gap equation. The numerical solution is presented in the respective sections dedicated to the analysis of these mesons.  

\subsection{The Gap Equation}
\begin{figure}[t!]
   \vspace{-3cm}
   \centering
    \includegraphics[scale=0.4,angle=0]{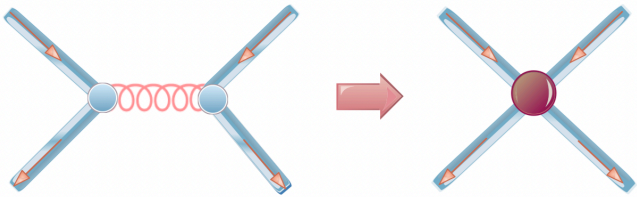}
    \vspace{-4cm}
    \caption{Diagrammatic representation of the CI, employing the simplified model of the gluon propagator in Eq.~(\ref{eqn:contact_interaction}).}
    \label{fig:ci}
\end{figure}

The starting point for our study is the dressed-quark propagator for a quark of flavor $f$,
which is obtained by solving the gap equation,
\begin{eqnarray}
 S(p)^{-1} &=& i\gamma\cdot p + m_{f} + \Sigma(p) \;,\nn \\
\Sigma(p) &=& \frac{4}{3} \int \! \frac{d^4q}{(2\pi)^4} g^2 D_{\mu\nu}(p-q)
\gamma_\mu S(q) \Gamma_\nu(q,p) ,\; \label{gendse}
\end{eqnarray}
where $m_f$ is the Lagrangian current-quark mass, $D_{\mu\nu}(p)$ is
the gluon propagator and $\Gamma_\nu(q,p)$ is the quark-gluon vertex. 
It is a well-established fact by now that the Landau gauge gluon propagator saturates in the infrared and a large effective mass scale is generated for the gluon, see for example~\cite{Boucaud:2011ug,Ayala:2012pb,Bashir:2013zha,Binosi:2016nme,Deur:2016tte,Rodriguez-Quintero:2018wma}. It also leads to the saturation of the effective strong coupling at large distances. This modern understanding of infrared QCD
forms the defining ideas of the CI proposed in~\cite{GutierrezGuerrero:2010md}.
We assume that the quarks interact, not through massless vector-boson exchange
but via a CI. Thus the gluon propagator no longer runs with a momentum scale but is frozen into a CI in keeping with the infrared properties of QCD, see Fig. \ref{fig:ci}.
Thus 
\begin{eqnarray}
\label{eqn:contact_interaction}
g^{2}D_{\mu \nu}(k)&=&4\pi\hat{\alpha}_{\mathrm{IR}}\delta_{\mu \nu} 
\end{eqnarray}
\noindent  where $\hat{\alpha}_{\mathrm{IR}}=\alpha_{\mathrm{IR}}/m_g^2$.
The scale $m_g$ is for dimensional reasons and is interpreted as the infrared gluon mass scale generated dynamically within QCD \cite{ Bowman:2004jm, Gutierrez-Guerrero:2010waf,Gutierrez-Guerrero:2019uwa}. We take currently accepted value $m_g=500\,\MeV$~\cite{Boucaud:2011ug,Aguilar:2017dco,Binosi:2017rwj,Gao:2017uox}. It is clear that in the CI gap equation, the effective coupling which appears is $\hat{\alpha}_{\mathrm{IR}}$ instead of $\alpha_{\mathrm{IR}}. \hspace{-2mm}$ 
 We choose $\alpha_{\rm IR}/\pi$ to be $0.36$ so that $\hat{\alpha}_{\mathrm{IR}}$ has exactly the same value as in all related previous works~\cite{Gutierrez-Guerrero:2010waf,Gutierrez-Guerrero:2019uwa,Gutierrez-Guerrero:2021rsx,Yin:2019bxe}.
The interaction vertex is bare, i.e.,
$\Gamma_\nu(q,p)=\gamma_\nu$.

This constitutes an algebraically simple but useful and predictive 
rainbow-ladder truncation of the SDE of the quark propagator whose
solution can readily be written as follows:
 \bea\label{DynamicalM}
 S(q,M_f) &\equiv &-i\gamma\cdot q \; \sigma_{V}(q,M_f)+\sigma_{S}(q,M_f)\,,
 \eea
 with
 \bea
 \sigma_{V}(q,M_f)=\frac{1}{q^{2}+M_f^{2}}\, , \sigma_{S}(q,M_f)= M_f
 \, \sigma_{V}(q,M_f) \,, 
 \eea
where $M_f$, for the CI, is momentum-independent dynamically generated dressed quark mass determined by
\begin{equation}
M_f = m_f + M_f\frac{4\hat{\alpha}_{\rm IR}}{3\pi}
\int_0^\infty \!ds \, s\, \frac{1}{s+M_f^2}\,\,. \label{gap-2}
\end{equation}
Our regularization procedure
 follows Ref.\,\cite{Ebert:1996vx}:
\begin{eqnarray}
\nonumber \frac{1}{s+M_f^2} & = & \int_0^\infty d\tau\,{\rm
e}^{-\tau (s+M_f^2)} \rightarrow \int_{\tau_{\rm UV}^2}^{\tau_{\rm
IR}^2} d\tau\,{\rm e}^{-\tau (s+M_f^2)}
\label{RegC}\\
& = & \frac{{\rm e}^{- (s+M_f^2)\tau_{\rm UV}^2}-e^{-(s+M_f^2)
\tau_{\rm IR}^2}}{s+M_f^2} \,, \label{ExplicitRS}
\end{eqnarray}
where $\tau_{\rm IR,UV}$ are, respectively, infrared and
ultraviolet regulators.  It is apparent from
Eq.\,(\ref{ExplicitRS}) that a finite value of $\tau_{\rm
IR}\equiv 1/\Lambda_{\rm IR}$ implements confinement by ensuring the
absence of quark production thresholds. Since Eq.\,(\ref{gap-2})
does not define a renormalisable theory, $\Lambda_{\rm
UV}\equiv 1/\tau_{\rm UV}$ cannot be removed but instead plays a
dynamical role, setting the scale of all mass dimensioned quantities.
Using Eq.\,\eqref{RegC}, the gap equation becomes
 \bea \label{eq:gapeq}
  M_f = m_f + M_f \frac{4 \hat{\alpha}_{\rm IR}}{3 \pi }
  {\cal C}(M_f^2) \;,
 \eea
 where
 \bea
  \frac{{\cal C}(M^2)}{M^2} = \Gamma(-1,M^2 \tau_{\rm UV}^2) -
  \Gamma(-1,M^2 \tau_{\rm IR}^2) \; 
 \eea
 and $\Gamma(\alpha,x)$ is the incomplete gamma-function.
 

 \begin{table}[t!]
 \caption{ \label{parameters} 
 Ultraviolet regulator and coupling constant for different combinations of quarks in PS mesons.  $\hat{\alpha}_{\mathrm {IR}}=\hat{\alpha}_{\mathrm{IRL}}/Z_H$, where $\hat{\alpha}_{\mathrm {IRL}}=4.57$ is extracted from the best-fit to data as explained in Ref.~\cite{Raya:2017ggu}.  $\Lambda_{\rm IR} = 0.24$ GeV is a fixed parameter.} 
\begin{center}
\label{parameters1}
\begin{tabular}{@{\extracolsep{0.0 cm}} || l | c | c | c ||}
\hline \hline
 \, quarks \, &\,  $Z_{H}$ \, &\,  $\Lambda_{\mathrm {UV}}\,[\GeV] $ \,  &\,  $\hat{\alpha}_{\mathrm {IR}}$
 \\
 \hline
 \rule{0ex}{2.5ex}
$\, \tu,\td,\ts$ & 1 & 0.905 & \, 
 4.57\,  \\ 
\rule{0ex}{2.5ex}
$\, \tc,\tu,\ts$ & \, 3.034 \, & 1.322 & 1.50 \\ 
\rule{0ex}{2.5ex}
$\, \tc$     &  13.122 & 2.305 & 0.35 \\
\rule{0ex}{2.5ex}
$\, \tb,\tu$ & 11.273 & 3.222 & 0.41 \\
\rule{0ex}{2.5ex}
 $\, \tb,\ts$ & 17.537 & 3.574 & 0.26 \\
\rule{0ex}{2.5ex}
$\, \tb,\tc$   & 30.537 & 3.886 & 0.15 \\
\rule{0ex}{2.5ex}
$\, \tb$     & 129.513 & 7.159 & \, 0.035 \,  \\
\hline \hline
\end{tabular}
\end{center}
\end{table}
 We report results for PS mesons  using the parameter values listed in Tables~\ref{parameters},~\ref{table-M}, whose variation with quark mass was dubbed as 
 {\em heavy parameters} in Ref.~\cite{Gutierrez-Guerrero:2019uwa}. 
 In this approach, the coupling constant and the ultraviolet regulator vary as a function of the quark mass. This behavior was first suggested  in Ref.~\cite{Bedolla:2015mpa} and later adopted in several subsequent works~\cite{Bedolla:2016yxq,Raya:2017ggu,Gutierrez-Guerrero:2019uwa,Yin:2019bxe,Yin:2021uom}. Table~\ref{table-M} presents the current quark masses $m_f$ used herein and the dynamically generated dressed masses $M_f$ of $\tu$, $\ts$, $\tc$ and $\tb$ computed from the gap equation, Eq.~(\ref{eq:gapeq}).
 
\begin{table}[h!]
\caption{\label{table-M}
Current ($m_{f}$) and dressed masses
($M_{f}$) for quarks in GeV, required as an input for the BSE and the EFFs.}
\vspace{0.3cm}
\begin{tabular}{@{\extracolsep{0.0 cm}} || c | c | c | c || }
\hline 
\hline
 $m_{\tu}=0.007$ &$m_{\ts}=0.17$ & $m_{\tc}=1.08$ & $m_{\tb}=3.92$   \\
 \rule{0ex}{2.5ex}
 $ M_{\tu}=0.367$ \, & \, $  M_{\ts}=0.53$\; \, &\,   $  M_{\tc}=1.52$ \, &\,  $  M_{\tb}=4.75$   \\
 \hline
 \hline
\end{tabular}
\end{table}

A meson can consist of heavy ($Q$) or light ($q$) quarks. We present the study of all heavy ($Q\bar{Q}$), heavy-light ($Q\bar{q}$) and (review) light ($q\bar{q}$) mesons.
We commence by setting up the BSE for mesons by employing a kernel which is consistent with that of the gap equation to obey axial vector Ward-Takahashi identity and low energy Goldberger-Treiman relations, see Ref.~\cite{Gutierrez-Guerrero:2010waf} for details. 
The PS mesons are $J^{PC}=0^{-+}$ states while the S mesons are $J^{PC}=0^{++}$ states. 
The solution of the BSE yields BSAs whose general form depends not only on the spin and parity of the meson under consideration  but also on the interaction employed as explained in the next sub-section.

\subsection{Bethe Salpeter Equation}
\begin{figure}[b!]
   \centering
    \includegraphics[scale=0.5,angle=0]{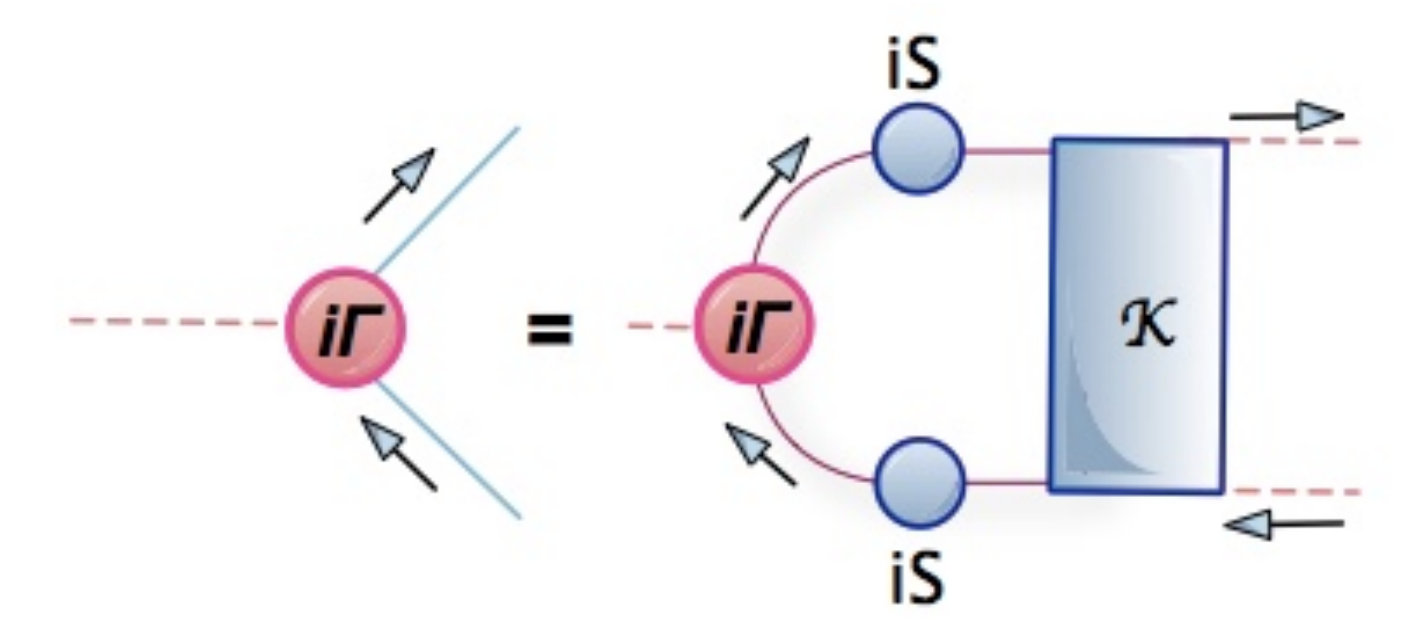}
    \caption{Diagrammatic representation of the BSE. Blue (solid) circles represent dressed quark propagators $S$, red (solid) circle is the meson BSA $\Gamma$ while the blue (solid) rectangle is the dressed-quark-antiquark scattering kernel ${\mathcal {K}}$.}
    \label{fig:BSEfig}
\end{figure}
  The relativistic bound-state problem for hadrons characterized by two valence-quarks may be studied using the
 homogeneous BSE whose diagrammatic representation can be seen in  Fig.~\ref{fig:BSEfig}. This equation is mathematically expressed  as~\cite{Salpeter:1951sz},
 \begin{equation}
[\Gamma(k;P)]_{tu} = \int \! \frac{d^4q}{(2\pi)^4} [\chi(q;P)]_{sr} {\mathcal K}_{tu}^{rs}(q,k;P)\,,
\label{genbse}
\end{equation}
where $[\Gamma(k;P)]_{tu}$ represents the bound-state's BSA and $\chi(q;P) = S(q+P)\Gamma S(q)$ is the BS wave-function; $r,s,t,u$ represent colour, flavor and spinor indices; and ${\mathcal K}$ is the relevant quark-antiquark scattering kernel. This equation possesses solutions on that discrete set of $P^2$-values for which bound-states exist.

A general decomposition of the BSA for the PS and the S mesons ($\fd\fu$)  in the CI has the following form
 \bea
\label{KaonBSA}
\Gamma_{\Meps}(P) &=& i \gamma_5 \,E_{\Meps}(P) + \frac{1}{2 M_R} \gamma_5 \gamma\cdot P \, F_{\Meps}(P) \nn \,,
\label{BSAs-Mesons} \\
 \Gamma_{\Ms}(P)&=&I_{D}\, E_{\Ms}(P) \,.
\eea
 Note that $E_i(P)$ and $F_i(P)$ with $i\in\{PS, S\}$ are known as the BSAs of the meson under consideration, $P$ is its total momentum,  $I_D$ is the identity matrix and  $M_R = M_{\fd} M_{\fu}/[M_{\fd} + M_{\fu}]$ is the reduced mass of the system. 
 Eq.~(\ref{genbse}) has a solution when $P^2=-M_{M}^2$ with $M_M$ being the meson mass. 
 After this initial and required set up of the gap equation and the BSE,   
 we now turn our attention to the description of the EFFs of mesons.


\section{\label{sec:eff} Electromagnetic Form Factors}

The EFFs provide crucial information on the internal structure of mesons. At low momenta, EFFs allow us to unravel the complexities of non-perturbative QCD, i.e., confinement, DCSB and the fully dressed quarks. At high energies, we expect to confirm the validity of asymptotic QCD for its realistic models while at intermediate energies, we observe a smooth transition from one facet of strong interactions to the other, all in one single experiment if we are able to chart out a wide range of momentum transfer squared $Q^2$ without breaking up the mesons under study. While there are plenty of studies on the pion EFFs, only a few are found about heavy-quarkonia and practically none on heavy-light mesons.
The process involves an incident photon which probes mesons, interacting with the electrically charged quarks making up these two-particles bound states. Therefore, it is natural to start this section by looking at the the structure of the quark-photon vertex within the CI.

\subsection{The Quark-Photon Vertex}
The quark-photon vertex, denoted by   $\Gamma_{\mu}^{\gamma}(k_+,k_-,M_{\fd})$, is related to the quark propagator through the following  vector Ward-Takahashi
identity:
 \bea
  i P_{\mu} \Gamma_{\mu}^{\gamma}(k_+,k_-,M_{\fd}) =
  S^{-1}(k_+,M_{\fd}) - S^{-1}(k_-,M_{\fd}) \,. \nonumber \\ \label{VWTI}
 \eea
 This identity is crucial for a sensible study of a bound-state's EFF. It is determined through the following inhomogeneous BSE,
 \bea
 && \hspace{-1.2cm} \Gamma_{\mu}^{\gamma}(Q,M_{\fd})= \nn \\
 && \gamma_{\mu} - \frac{16 \pi \hat{\alpha}_{\rm IR}}{3} 
  \int  \frac{d^4q}{(2 \pi)^4} \gamma_{\alpha} \chi_{\mu}(q_+,q,M_{\fd})
 \gamma_{\alpha} \, ,\label{eqvertex}
 \eea
where $\chi_{\mu}(q_+,q,M_{\fd})  =  S(q+P,M_{\fd}) \Gamma_{\mu}(Q)S(q,M_{\fd})$.
Owing to the momentum-independent nature of the interaction
kernel, the general form of the solution is
  \bea
 && \hspace{-4mm} \Gamma_{\mu}^{\gamma}(Q,M_{\fd})= \gamma_{\mu}^{L}(Q)P_{L}(Q^{2},M_{\fd}) +
 \gamma_{\mu}^{T}(Q)P_{T}(Q^{2},M_{\fd}), \nonumber \\
 \eea
 where $\gamma_{\mu}^{L} + \gamma_{\mu}^{T} = \gamma_{\mu}$ and
  \bea \gamma_{\mu}^{T}(Q)=\gamma_{\mu}-\frac{ \gamma \cdot Q \;
 }{Q^{2}} \, Q_{\mu}  \,.
 \eea
 Inserting this general form into Eq.~(\ref{eqvertex}), one readily
 obtains (on simplifying notation)
 \bea
 \hspace{-2mm} P_{L} =1 \,, \quad 
 P_T= \frac{1}{1+K_\gamma(Q^2,M_{\fd})} , \label{PTQ2}
 \eea
\vspace{-0.5cm}
\begin{figure}[htbp]
\centering
\includegraphics[width=9cm]{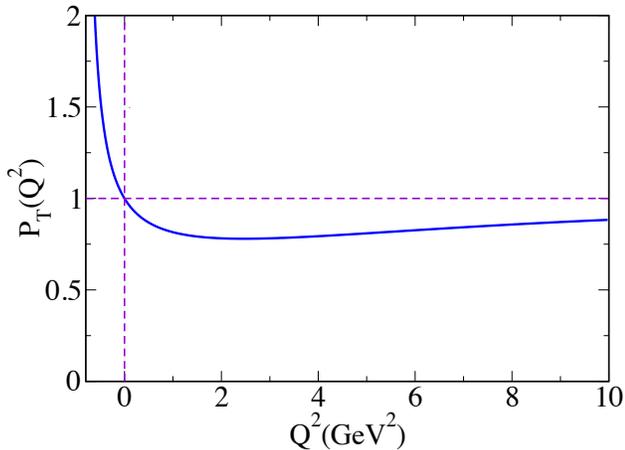}
\caption{Dressing function of the transverse quark-photon vertex, $P_T(Q^2)$, in~\protect{Eq.~(\ref{PTQ2}).}}
\label{fig:rhopole}
\end{figure}


\noindent with
\begin{align}
 K_\gamma(Q^2,&M_{\fd}) = \frac{4 \hat{\alpha}_{\rm
IR}}{3\pi} 
\int_0^1d\alpha\, \alpha(1-\alpha) Q^2\,\bar{\mathcal{C}}_1(\omega)
\,,
\end{align}
where
\bea
\bar{\cal C}_1(z) = - \frac{d}{dz}{\cal
C}(z)= \Gamma(0,z\,  \tau_{\rm UV}^2)-\Gamma(0,z \,
\tau_{\rm IR}^2)\,
\eea
and
\bea
 \omega&=&\omega(M_{\fd}^2,\alpha,Q^2) 
 =M_{\fd}^2 + \alpha(1-\alpha) Q^2 \,.
\eea
One can clearly observe from Fig.~\ref{fig:rhopole} that $P_{T}(Q^{2})
\rightarrow 1$ when $Q^2 \rightarrow \infty$, yielding the
perturbative bare vertex $\gamma_{\mu}$ as expected.
This quark-photon vertex provides us with the required electromagnetic interaction capable of probing the EFFs of mesons through a triangle diagram which keeps the identity of the meson bound state intact.



\subsection{The Triangle Diagram}

\begin{figure}[b]
\centerline{
\includegraphics[scale=0.25,angle=0]{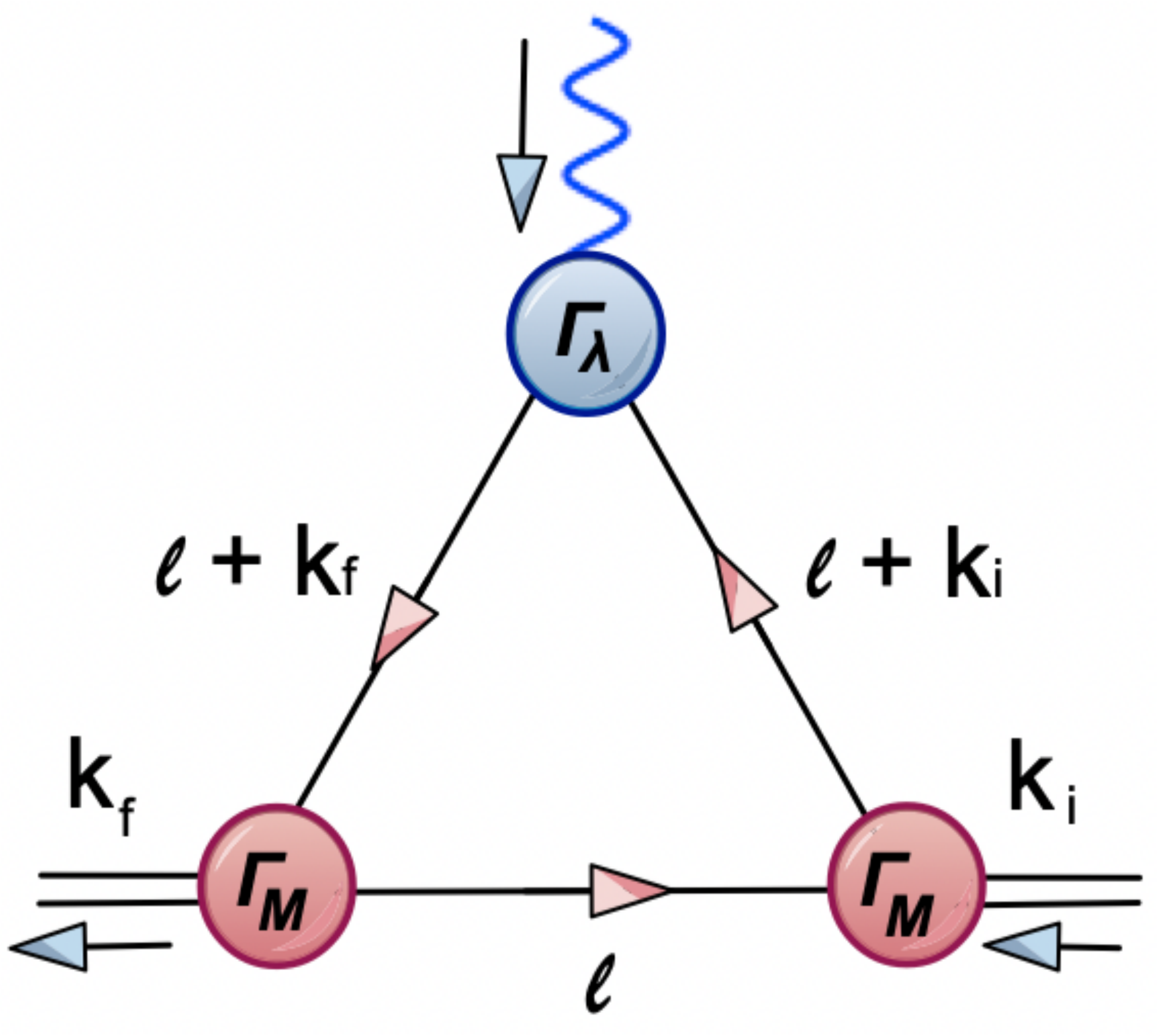}
}
    \caption{\label{vertex-1} The triangle diagram for the impulse approximation to the $M\gamma M$ vertex.}
\end{figure}

Let us start from the general considerations for the  electromagnetic
interactions of mesons. In the impulse approximation, the
$M \gamma M$ vertex, which describes the interaction between a meson ($\fd\fu$) and a photon, reads
\bea \Lambda^{M,\fd}&=&N_c\int \frac{d^{4}\ell}{(2\pi)^{4}}
 {\rm Tr}\;\mathcal{G}^{M,\fd} \,,
 \eea
where
 \bea
 \nonumber\mathcal{G}^{M,\fd}&=& 
  \, i\Gamma_{M}(k_{f}) \, S(\ell+k_{i},M_{\fd}) \, i\Gamma_{\lambda}(Q,M_{\fd})\\
  \nn &\times& 
  \, S(\ell+k_{f},M_{\fd})
 i\bar{\Gamma}_{M}(-k_{i}) \, S(\ell,M_{\fu}) \,. \label{General-FF}
 \eea
The notation assumes that it is the quark $\fd$ which interacts with the photon while the antiquark $\fu$ remains a spectator. We  define 
$\Lambda^{M,\fu}$ similarly. 
 Furthermore, we denote the incoming photon momentum by $Q$ while the incoming and outgoing momenta of $M$ by:
$k_{i}=k-Q/2$ and $k_f=k+Q/2$, respectively.
The assignments of momenta are shown in the triangle diagram of Fig.~\ref{vertex-1}.

$\Lambda^{M,\f}$ corresponds to the EFFs of different mesons under study. The contribution from the interaction of the photon with quark $\fd$ can be represented as $F^{M,\fd} (Q^2)$ (stemming from $\Lambda^{M,\fd}$) while the contribution arising from its interaction with quark $\fu$ can be represented as $F^{M,\fu} (Q^2)$ (coming from $\Lambda^{M,\fu}$). The total form factor $F^M (Q^2)$ is defined as follows~\cite{Hutauruk:2016sug}:
\begin{equation}\label{eqn:TotalMesonFF}
F^M (Q^2) = e_{\fd} F^{M,\fd} (Q^2) + e_{\fu} F^{M,\fu} (Q^2)\,,
\end{equation}
where $e_{\fd}$ and $e_{\fu}$ are the quark and the antiquark electric charges, respectively~\footnote{For neutral mesons composed of same flavored quarks, the total EFF is simply $F^{M} = F^{M,\fd}$.}. Both for PS  and S mesons, $F^{M,f_1}$ is straightforwardly related to $\Lambda^{M,\fd}$:
\begin{eqnarray}
\Lambda^{\Ms,\fd}=-2k_\lambda F^{\Ms,\fd} \,, \, \Lambda^{\Meps,\fd}=-2k_\lambda F^{\Meps,\fd} \,.
\end{eqnarray}

\noindent
All information necessary for
the calculation of the EFFs is now complete. We can employ numerical values of the parameters listed in Tables~\ref{parameters} and~\ref{table-M} and proceed to compute the EFFs.
Our evaluated analytical expressions and numerical results for PS and S mesons occupy the details of the next two sections. Keeping in mind that the pairs of (PS,~S) mesons can be considered as parity partners, we embark upon their treatment in the following sections in that order.  

\section{\label{PS-S-FF} Pseudoscalar Mesons}

\begin{table}[b!]
\caption{\label{par-AllFF} Calculated values for the BSAs and masses for PS mesons in CI model
 using the parameters in Tables~\ref{parameters} 
and~\ref{table-M} (compare the parameters with  the ones in Ref.~\cite{Gutierrez-Guerrero:2019uwa}).} 
\begin{tabular}{@{\extracolsep{0.1 cm}} || c | ccc | c | c || }
\toprule
 \rule{0ex}{2.5ex}
 &  Mass[GeV]  & $E_{\Meps}$ & $F_{\Meps}$ & $m_{PS}^{\rm exp}$[GeV] & error [\%]  \\ 
 \hline
 \rule{0ex}{2.5ex}
 $\tu\bar\td$ \,& 0.139 & 3.59 & 0.47 & 0.139 & 0.008  \\
 \rule{0ex}{2.5ex}
 $\tu\bar\ts$ \,& 0.499 & 3.81 & 0.59  & 0.493 & 1.162  \\
\rule{0ex}{2.5ex}
$\ts\bar{\ts}$ \,& 0.701 & 4.04 & 0.75  & --- & ---  \\
\rule{0ex}{2.5ex}
$\tc\bar{\tu}$ \,& 1.855 & 3.03 & 0.37  & 1.864 & 0.494  \\
\rule{0ex}{2.5ex}
$\tc\bar{\ts}$ \,& 1.945 & 3.24 & 0.51 & 1.986 & 1.183 \\
\rule{0ex}{2.5ex}
$\tu\bar{\tb}$ \,& 5.082 & 3.72 & 0.21 & 5.279 & 3.735  \\
\rule{0ex}{2.5ex}
$\ts\bar{\tb}$ \,& 5.281 & 2.85 & 0.21  & 5.366 & 1.586  \\
\rule{0ex}{2.5ex}
$\tc\bar{\tb}$ \,&  6.138 & 2.58 & 0.39 & 6.274 & 2.166  \\
\rule{0ex}{2.5ex}
$\tc\bar{\tc}$ \,& 2.952 & 2.15 & 0.40 & 2.983 & 1.053  \\
\rule{0ex}{2.5ex}
$\tb\bar{\tb}$ \,& 9.280 & 2.04 & 0.39 & 9.398 & 1.262  \\
\hline
\hline
\end{tabular}
\end{table}
%

\begin{figure*}[t!]
\begin{tabular}{@{\extracolsep{-2.3 cm}}c}
 \renewcommand{\arraystretch}{-1.6} %
 \hspace{-1cm}
 \includegraphics[scale=0.7]{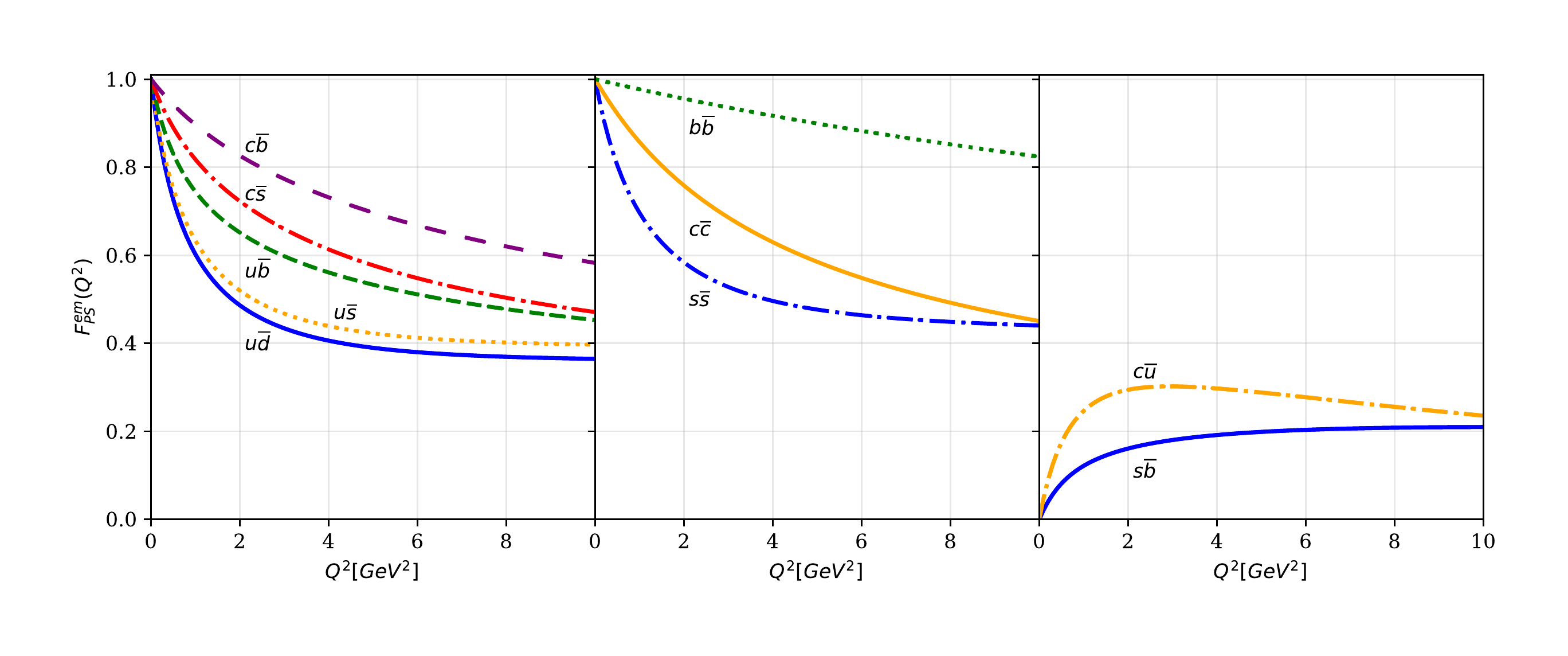}
\end{tabular}
\vspace{-1cm}
\caption{\label{plotPS} EFFs of PS mesons in a CI model. Left panel: electrically charged mesons composed of quarks of different flavors. Central panel: quarkonia including a hypothetical ground state {\em strangeonium} ($\ts\bar \ts$). Right panel: electrically neutral mesons composed of quarks of different flavors.}
\end{figure*}

We start with a detailed discussion and results on the ground state PS mesons. These are negative parity, zero angular momentum $0^{-+}$ states and occupy a special role in hadron physics. Simultaneously these are the simplest bound states of a quark and antiquark and also emerge as Goldstone bosons associated with DCSB. 
Pions are the lightest hadrons
  and are produced copiously in collider machines at all energies. The pion cloud effect substantially contributes to several static and dynamical hadron properties. Therefore, understanding their internal structure has been of great interest both for experimenters and theoreticians. 
The study of PS mesons is crucial in understanding the capabilities and limitations of the CI model employed in this work to reproduce and predict phenomenological results. 
Being the Goldstone bosons associated with DCSB, their analysis requires care in treating the associated subtleties. 
From Eqs.~(\ref{BSAs-Mesons}), we can see that the BSA of PS mesons is the only one to be composed of two terms, necessary to ensure the axial vector Ward-Takahashi identity and the Goldberger-Treiman relations are exactly satisfied.  
In this article, we extend and expand the work presented in~\cite{Gutierrez-Guerrero:2010waf,Bedolla:2016yxq,Raya:2017ggu} and compute the EFFs of a larger number of PS mesons composed of $qq, qQ$ and $QQ$ quarks. With straightforward algebraic manipulations:
\bea \nn
&& \hspace{-8mm} F^{\Meps,\fd}= 
P_T(Q^2)\bigg[ 
E_{\Meps}^2 T^{\Meps}_{EE}(Q^2) +
E_{\Meps} F_{\Meps} T^{\Meps}_{EF}(Q^2)\\
&& \hspace{+6.6mm}  +
F_{\Meps}^2 T^{\Meps}_{FF}(Q^2) \bigg]\,,
\eea
where
\begin{align}
\nn T^{\Meps}_{EE}(Q^2) & = \frac{3}{4\pi^2}\bigg[ \int_0^1 d\alpha\, \overline{\mathcal{C}}_1(\omega_1) \\
\nn &+ 
2\int_0^1 d\alpha\, d\beta \, \alpha \, \mathcal{A}^{\Meps}_{EE} \, \overline{\mathcal{C}}_2(\omega_2) \bigg]\,,\\
\nn T^{\Meps}_{EF}(Q^2) & 
= -\frac{3}{2\pi^2} \frac{1}{M_R}\int_0^1 d\alpha \, d\beta \, \alpha \bigg[\mathcal{A}_{EF}^{(1)} \, \overline{\mathcal{C}}_1(\omega_2)\\
\nn &+  
(\mathcal{A}_{EF}^{(2)}-\omega_2 \mathcal{A}_{EF}^{(1)}) \, \overline{\mathcal{C}}_2(\omega_2) \bigg]\,,\\
\nn T^{\Meps}_{FF}(Q^2) & =  \frac{3}{4\pi^2} \frac{1}{M_R^2}
\int_0^1 d\alpha \, d\beta \, \alpha \bigg[ \mathcal{A}_{FF}^{(1)} \, \overline{\mathcal{C}}_1(\omega_2)  \\
\nn &+
 (\mathcal{A}_{FF}^{(2)} -\omega_2 \mathcal{A}_{FF}^{(1)}) \, \overline{\mathcal{C}}_2(\omega_2)  \bigg]\,,
\end{align}
and
\begin{align}
\nn \omega_1 &= \omega_1(M_{\fd},\alpha,Q^2) = M_{\fd}^2+\alpha\, Q^2(1-\alpha) \, ,\\
\omega_2&=\omega_2(M_{\fd},M_{\fu},\alpha,\beta,M_M) = \alpha \,  M_{\fd}^2 +(1-\alpha)M_{\fu}^2 \nn \\  &  \nn  -\alpha(1-\alpha)\, M^2_{M} +\alpha^2 \, \beta \,  (1-\beta) \, Q^2\, , \nn \\
    \bar{\mathcal{C}}_2(z)&=
(\exp(-z\, \tau_{\rm uv})-\exp(-z\,\tau_{\rm ir}))/(2z)\,.
\end{align}
The coefficients ${\cal A}_i$ are given by the following expressions:
\bea
\nn
&& \hspace{-4mm} \mathcal{A}^{\Meps}_{EE} = \alpha (M_{\fd}^2 \hspace{-1mm} + \hspace{-1mm} M^2_{M})+2(1-\alpha) M_{\fd} M_{\fu} + (\alpha-2)M_{\fu}^2 \,, \nn \\
\nn && \hspace{-4mm}  \mathcal{A}_{EF}^{(1)} 
= M_{\fd} +M_{\fu}\,, \\
\nn && \hspace{-4mm}  \mathcal{A}_{EF}^{(2)} =  
2 M_{\fd}^2 M_{\fu}-\alpha M_{\fd} (4 (\alpha-1) M_{M}^2+\alpha Q^2)\\
\nn && \hspace{+4.2mm} +M_{\fu} (2 (\alpha-1)^2 M_{M}^2 
+\alpha Q^2 (2 \alpha (\beta-1) \beta \hspace{-0.5mm} + \hspace{-0.5mm}\alpha-1))\,,
\\
\nn && \hspace{-4mm}  \mathcal{A}_{FF}^{(1)}=  (3\alpha -2) M_M^2 +\alpha Q^2 \, , \\
\nn && \hspace{-4mm}  \mathcal{A}_{FF}^{(2)}=  2 \alpha ((\alpha-1)^2 M_M^4\\
\nn && \hspace{+4.2mm} + \, \alpha M_M^2 Q^2 (3 \alpha \beta^2-3 \alpha \beta+\alpha-2 \beta^2+2 \beta-1)) \nonumber \\
\nn && \hspace{+4.2mm} + 2 \alpha M_M^2 M_{\fd}^2-2 M_{\fd} M_{\fu} (2 (\alpha-1) M_M^2+\alpha Q^2) \,.
\eea 
The resulting EFFs for charged as well as neutral mesons are shown in Fig.~\ref{plotPS}.
For the practical utility and intuitive understanding of their low and large $Q^2$ behavior, we perform an 
interpolation for PS mesons EFF in the range $Q^2\in [ 0,8M_{M}^2 ]$. We adopt the following  functional form: 
\begin{align}
F^{PS}(Q^2) = \frac{e_{M} + a_{PS}\, Q^2 +b_{PS}\,Q^4}{1+c_{PS}\, Q^2 + d_{PS}\, Q^4}\,,
\label{EqfitPS}
\end{align}
where $e_M:=F^{PS}(Q^2=0)$ is the electric charge of the meson and $a_{PS},b_{PS},c_{PS},d_{PS}$ are the fitted coefficients. The best fit corresponds to the values listed in Table \ref{fitParaPS}.
The fit of Eq.~(\ref{EqfitPS}) resonates with our observation that
the EFFs of PS mesons tend to constant values for large $Q^2$ when it becomes by far the largest energy scale in the problem. It is a straightforward consequence of CI treatment, and it is characteristic of a point-like interaction which leads to harder EFF. However, the heavy as well as heavy-light mesons approach a constant value much slower than the light ones. This comparative large $Q^2$ behavior of EFFs owes itself to the fact that $Q^2$ becomes larger than all other energy scales at much higher values.

\begin{table}[hbt]
\caption{\label{fitParaPS}Parameters from the fit of Eq.(\ref{EqfitPS}) for PS mesons}
\begin{center}
\begin{tabular}{@{\extracolsep{0.0 cm}} || c | c | c | c | c || }
\hline
\hline
& $a_{PS}$ & $b_{PS}$ & $c_{PS}$ & $d_{PS}$  \\ 
\hline
\rule{0ex}{2.5ex}
 $\tu\bar{\td}$ \,& \,\, 0.330\,\,   &\,\,   0.029\,\,   & \, \, 1.190\,\,   &\,\,   0.068\,\,   \\
\rule{0ex}{2.5ex}
$ \tu\bar{\ts}$ \,& 0.335 & 0.029 & 1.092 & 0.065  \\
\rule{0ex}{2.5ex}
$\ts\bar{\ts}$ \,& 0.328 & 0.040 & 0.874 & 0.092  \\
\rule{0ex}{2.5ex}
$\tc\bar{\tu}$ \,& 0.616 & $-$0.001 & 1.370 & 0.109 \\
\rule{0ex}{2.5ex}
$\tc\bar{\ts}$ \,& 0.615 & 0.028 & 0.897 & 0.111  \\
\rule{0ex}{2.5ex}
$\tu\bar{\tb}$ \,& 1.143 & 0.033 & 1.921 & 0.146  \\
\rule{0ex}{2.5ex}
$\ts\bar{\tb}$ \,& 0.218 & 0.000 & 0.840 & 0.009 \\
\rule{0ex}{2.5ex}
$\tc\bar{\tb}$ \,& 0.333 & 0.003 & 0.493 & 0.021  \\
\rule{0ex}{2.5ex}
$\tc\bar{\tc}$ \,& 1.778 & 0.057 & 1.994 & 0.334  \\
\rule{0ex}{2.5ex}
$\tb\bar{\tb}$ \,& 0.099 & 0.000 & 0.127 & 0.002   \\
\hline \hline
\end{tabular}
\end{center}
\end{table}

 The behavior of the form factors at the other extreme, $Q^2 \simeq 0$ allows us to extract charge radii: 
\begin{equation}
 r_{M}^2 =
-6\left.\frac{\mathrm{d}F_{M}(Q^{2})}{\mathrm{d}Q^{2}}\right|_{Q^{2}=0}
\,.
\end{equation}

\begin{figure}[t!]
     \centerline{
     \includegraphics[scale=0.82,angle=0]{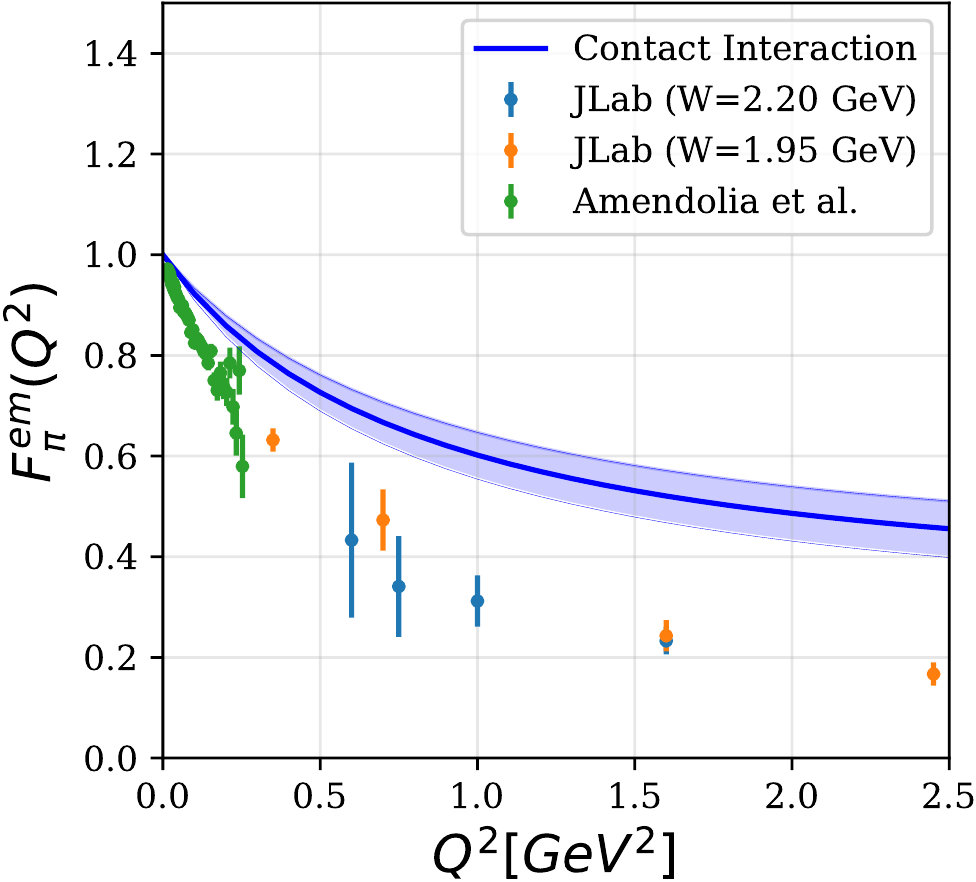}
       }
       \caption{EFF for $\pi$-meson. The central curve is obtained using the $\tau_{\rm UV}$ value from the \tab{parameters1}. The filled band allows for a $5\%$ variation in the charge radius. Dots represent the experimental data from Refs.~\cite{AMENDOLIA1986168,JeffersonLabFpi:2000nlc,JeffersonLabFpi-2:2006ysh}.}
       \label{pionS}
\end{figure}

\noindent  
For $\tc\bar{\tu}$  and $\ts\bar{\tb}$ states, which are normalized to $F^{M}(0)=0$, we define $r_{M}^2$ with a positive sign in the above equation. The charge radii set the trend for the subsequent evolution of the form factors as a function of $Q^2$, specially for its small and intermediate values. Fig.~\ref{fig:PS_hierarchy} depicts charge radii for all the PS mesons studied, allowing for a 5\% variation around the central value. With this permitted spread in the charge radii, one can obtain a band for the $Q^2$ evolution of the EFFs. To avoid over-crowding, we have avoided depicting such a band for each EFF. However, Fig.~\ref{pionS} shows a representative plot for the pion permitting a 5\% variation in its charge radius in conjunction with the available experimental results. 

Finally we list the central values of all ground state PS mesons charge radii in Table~\ref{tab:chargeradiuspsu}, along with a direct comparison with available experimental observations, lattice results and the SDE findings. Moreover, we also report the transition charge radii of light PS mesons and flavorless neutral heavy PS mesons to two photons invoking the following analytical parameter fit~\cite{Ding:2018xwy}:
\begin{eqnarray}
 r^t_M = \frac{r_0}{1+ (M_M/m_t) \, {\rm ln}(1+M_M/m_t) } \,, \label{fit-cr}
\end{eqnarray}
where $r_0=0.67$ fm and $m_t=1.01$ GeV. It also yields reasonable results for the $\pi$ point (mass = 0.139 GeV) and the $K$ point (mass = 0.493 GeV) as they are made of light quarks. But we cannot expect it to serve exactly as it is for mesons with vastly off-balanced quark masses. However, if CI results were to follow this formula, we would only need to assign $r_0=0.458$ fm. The last row of Table~\ref{tab:chargeradiuspsu} lists the resulting values which we denote as $r^{t}_M(CI)$. Let us now summarize our findings and make explicit comparisons with related works:

\begin{figure}
    \centering
    \includegraphics[scale=0.44]{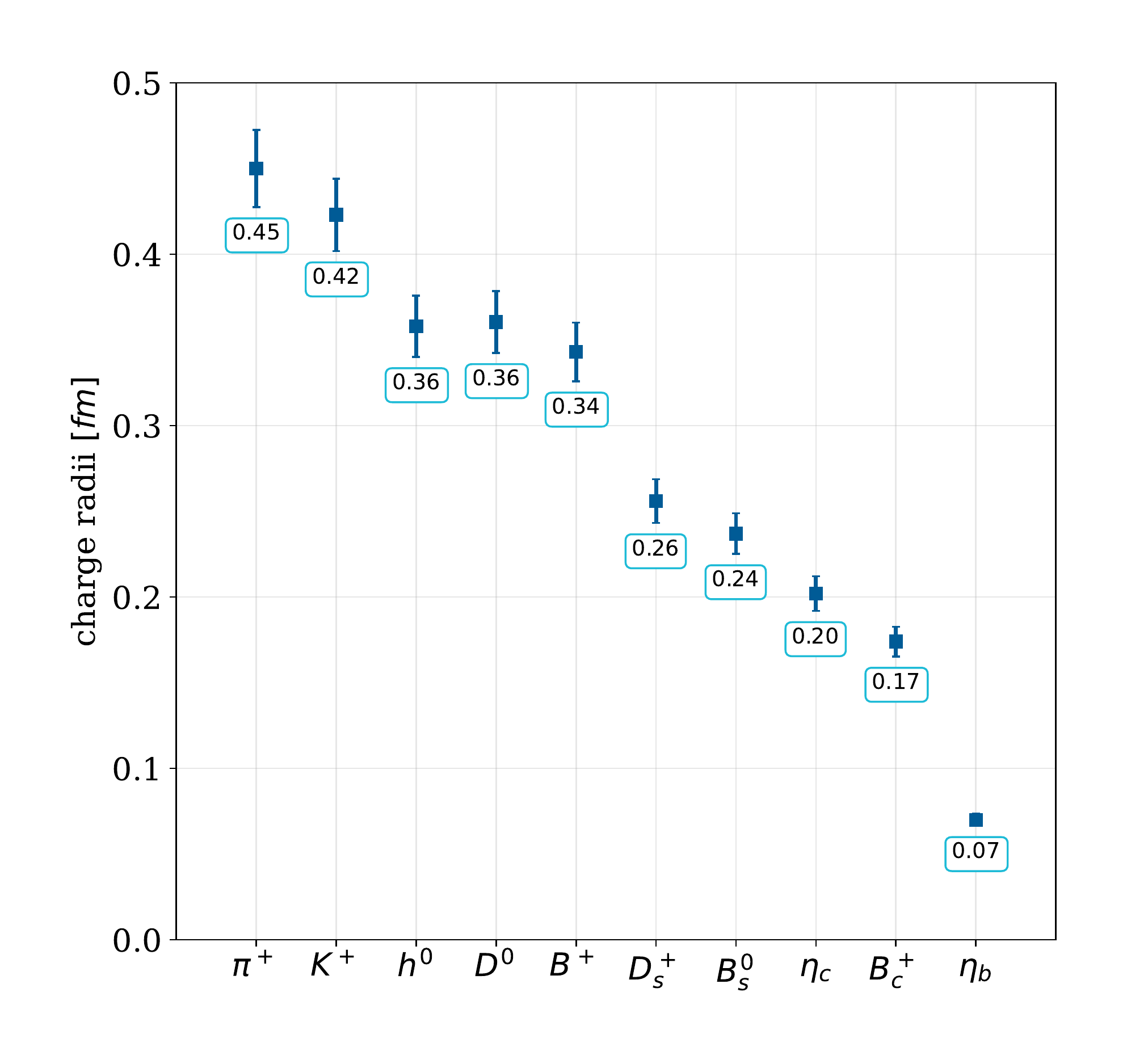}
    \caption{Charge radii of ground state PS mesons in the CI.}
    \label{fig:PS_hierarchy}
\end{figure}

\begin{figure}[t!]
     \centerline{\includegraphics[scale=0.82,angle=0]{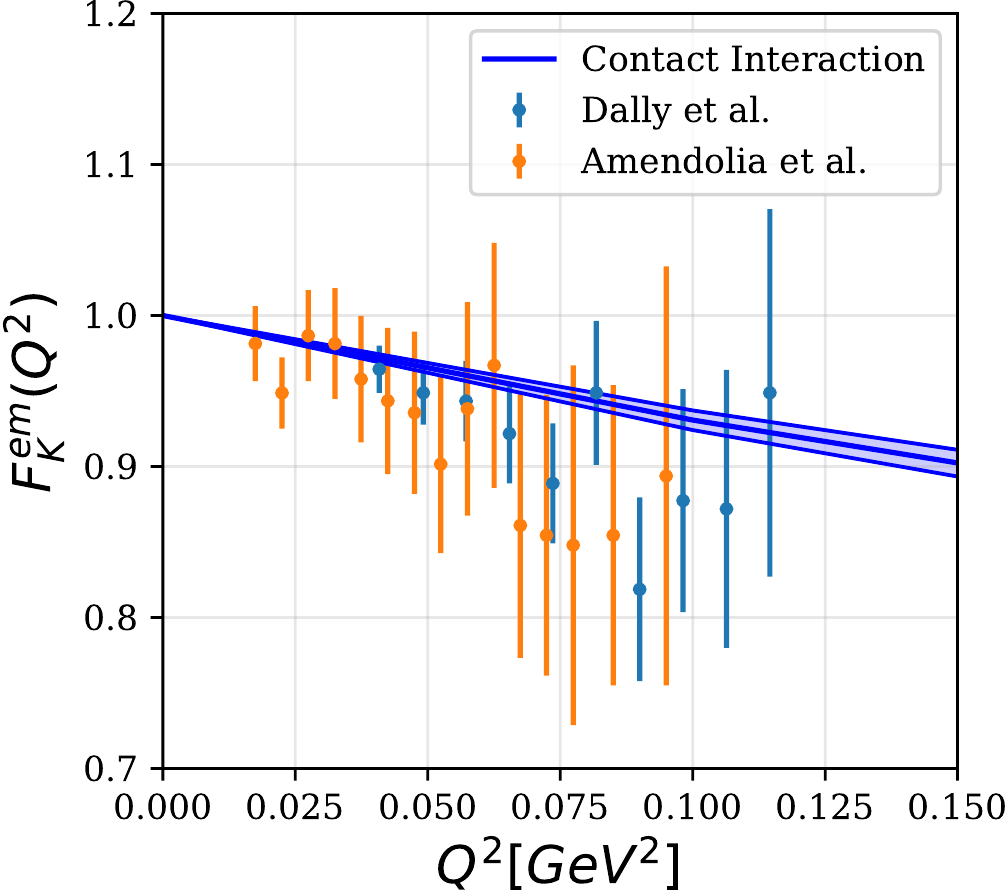}
       }
       \caption{EFF for $K$-meson. The central curve of the (blue) band is obtained by using the $\Lambda_{\rm UV}$ value from~\tab{parameters1}. The filled (blue) band allows for a $5\%$ variation in the charge radius. The experimental data is from Ref.~\cite{AMENDOLIA1986168}.}
       \label{kaonS}
\end{figure}
\begin{figure}[t!]
     \centerline{
     \includegraphics[scale=0.82,angle=0]{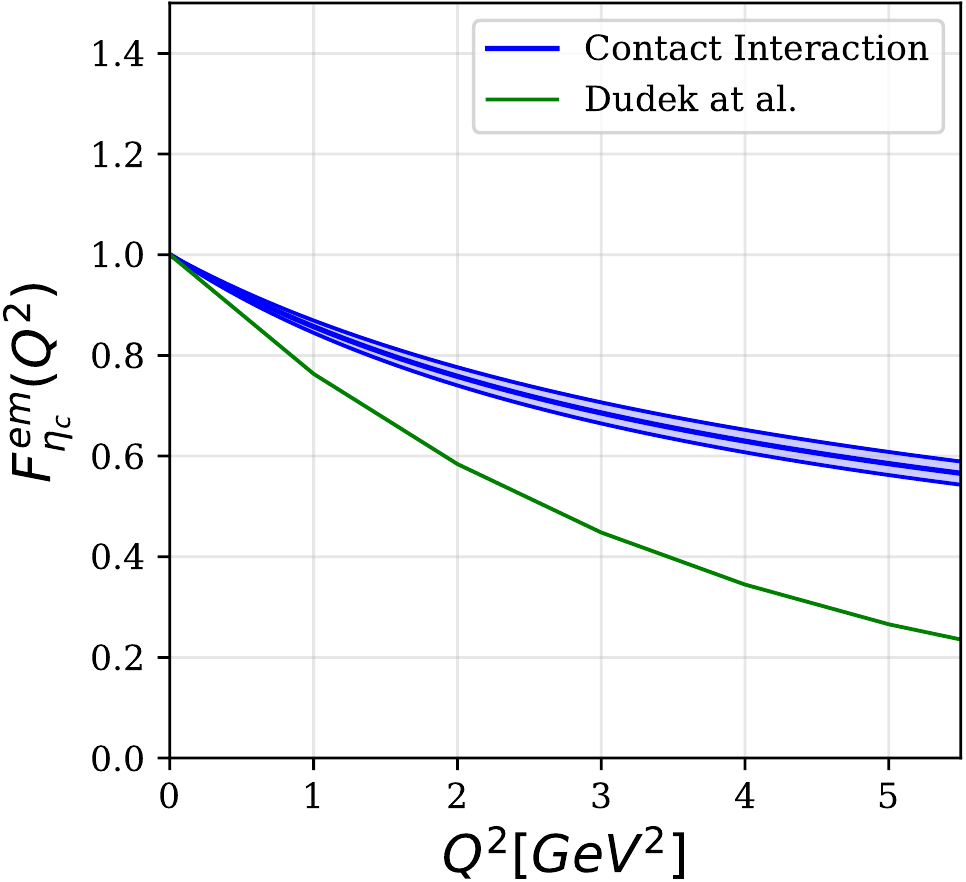}
       }
       \caption{EFF for $\eta_c$-meson. The lower (green) solid curve is the lattice result, Ref.~\cite{Dudek:2007zz}. The central curve of the (blue) band is obtained using the $\Lambda_{\rm UV}$ value from \tab{parameters1}. The filled (blue) band allows for a $5\%$ variation in the charge radius.}
       \label{etacS}
\end{figure}
\begin{table*}[t!]
\begin{center}
\caption{\label{tab:chargeradiuspsu} The charge radii of PS meson systems, calculated with the CI model, refined SDE studies, lattice QCD and extraction from data, in a hybrid model (HM), light-front framework (LFF) and a QCD potential model (PM). In the two rows after the experimental results, we also provide the best fit results for the transition charge radii of PS mesons to $\gamma \gamma^*$, Eq.~(\ref{fit-cr}) and the same fit adapted to the CI. All results are presented in fm.}
\begin{tabular}{@{\extracolsep{0.0 cm}} ||c|c|c|c|c|c|c|c|c|c|c||}
\hline\hline
\rule{0ex}{2.5ex}
&$\tu\bar{\td}$  & $\tu\bar{\ts}$ & $\ts\bar{\ts}$ & $\tc\bar{\tu}$ & $\tc\bar{\ts}$  &$\tu\bar{\tb}$ & $\ts\bar{\tb}$ & $\tc\bar{\tb}$ & $\tc\bar{\tc}$ & $\tb\bar{\tb}$ \\
\hline
 \rule{0ex}{2.5ex}
Our Result & 0.45 & 0.42  & \, 0.36\,   & \, 0.36\,  & \, 0.26\,  & \, 0.34\,  & \, 0.24\,  & \, 0.17\,  & \, 0.20\,  & \, 0.07\,  \\
 \rule{0ex}{2.5ex}
  SDE~\cite{Miramontes:2021exi,Bhagwat:2006pu} &  $0.676 \pm 0.002$ & $0.593 \pm 0.002$ &- &- &- &- &- &- & \, 0.24\,  & \, 0.09\,
   \\
\rule{0ex}{2.5ex}
  Lattice~\cite{Gao:2021xsm,Dudek:2006ej,Davies:2018zav} & \, $0.648 \pm 0.141$ \, & \, 0.566 {\rm (extracted)} \, &- &- &- &- &- &- & \, 0.25\,  & - \\
 \rule{0ex}{2.5ex}
  Exp.~\cite{ParticleDataGroup:2020ssz} &  $0.659 \pm 0.004$ & $0.560\pm 0.031$ &- &- &- &- &- &- &- &-  \\
  \rule{0ex}{2.5ex}
  $r^t_M$~\cite{Ding:2018xwy} &  $0.658$ & $0.568$ & - & - &- &- & - &- &\, 0.13\, &\, 0.03\, \\
   \rule{0ex}{2.5ex}
  $r^t_M({\rm CI})$ &  $0.45$ & $0.38$ & \, 0.33\, & - &- &- & - &- &\, 0.09\, &\, 0.02\,
 \\
\rule{0ex}{2.5ex}
  HM~\cite{Lombard:2000kw} & 0.66 & 0.65 & - & 0.47 & 0.50 & - & - & - & - & -
\\  
\rule{0ex}{2.5ex}
   LFF~\cite{Hwang:2001th} & 0.66 & 0.58 & - & 0.55 & 0.35 & 0.61 & 0.34 & 0.20 & - & -
\\   
\rule{0ex}{2.5ex}
PM~\cite{Das:2016rio} & - & - & - & 0.67 & 0.46 & 0.73 & 0.46 & - & -& -
\\ \hline\hline
\end{tabular}
\end{center}
\end{table*}

\begin{itemize}

\item 

As desired, pion EFF and its charge radius
agree with the first results employing the CI~\cite{Gutierrez-Guerrero:2010waf}. As an add-on, in this article we allow for a 5\% variation of the pion charge radius to see its effect on the evolution of the EFF as a function of $Q^2$, Fig.~\ref{pionS}. A small variation of the initial slope of the curve 
$Q^2 \simeq 0$ opens a noticeable spread for large $Q^2$ but keeps the qualitative and quantitative behaviour fully intact. 

\item In Fig.~\ref{kaonS}, we draw kaon EFF over the range of $Q^2$ values where (relatively poor) experimental observations are available. Although large error bars prevent us from 
commenting decisively on the validity of the CI but we expect it will yield harder results as compared to precise experimental measurements whenever these results will become available. Our reported value of its charge radius is an indication of this behavior. 

\item
As depicted in Table~\ref{tab:chargeradiuspsu}, pion and kaon charge radii~\cite{ParticleDataGroup:2020ssz,Gao:2021xsm,Dudek:2006ej,Davies:2018zav,Miramontes:2021exi,Bhagwat:2006pu} are known experimentally and through lattice and SDE studies. As CI EFFs come out to be harder than {\em full} QCD predictions, we expect our PS mesons charge radii to undershoot the exact results. This is precisely what we observe for the pion and the kaon. The percentage relative difference between the experimental value and our calculation for the pion charge radius is approximately $32\%$, while for the kaon charge radius is slightly less, $25\%$. Similar difference between the SDE and the CI results for heavy quarkonia is observed: For $\eta_c$, it is $20\%$ while for $\eta_b$ is is $22\%$, not too dissimilar. This comparatively analogous behavior augments our expectation that we will be in the same ballpark for the PS mesons whose charge radii are neither known experimentally as yet nor lattice offers any results.

\item
There are no experimental or lattice (to the best of our knowledge) results available for $\tc\bar{\tu}$, $\tc\bar{\ts}$, $\tu\bar{\tb}$, $\ts\bar{\tb}$ and $\tc\bar{\tb}$ mesons for comparison. However, the general trend of decreasing charge radii with increasing constituent quark mass seems reassuring, e.g., the following hierarchies are noticeable:
\begin{eqnarray}
&& r_{\tu\bar{\td}} > r_{\tu\bar{\ts}} > r_{\tc\bar{\tu}} > r_{\tu\bar{\tb}} \,, \nn \\
&& r_{\tu\bar{\ts}} > r_{\ts\bar{\ts}} > r_{\tc\bar{\ts}} > r_{\ts\bar{\tb}} \,, \nn \\
&& r_{\tc\bar{\tu}} > r_{\tc\bar{\ts}} > r_{\tc\bar{\tc}} > r_{\tc\bar{\tb}} \,, \nn \\
&& r_{\tu\bar{\tu}} > r_{\ts\bar{\ts}} > r_{\tc\bar{\tc}} > r_{\tb\bar{\tb}} \,.\nn 
\end{eqnarray}
We must emphasize that the CI is only a simple model. Refined QCD calculations are required to confirm or refute these findings.

\end{itemize}
This concludes our detailed analysis of all the ground state PS heavy ($Q\bar{Q}$), heavy-light ($Q\bar{q}$) as well as light ($q\bar{q}$) mesons. We now turn our attention to a similar analysis of the scalar mesons. 




\section{\label{S-FF} Scalar mesons}

\begin{figure}[tp!]`
       \centerline{
       \includegraphics[scale=0.3,angle=0]{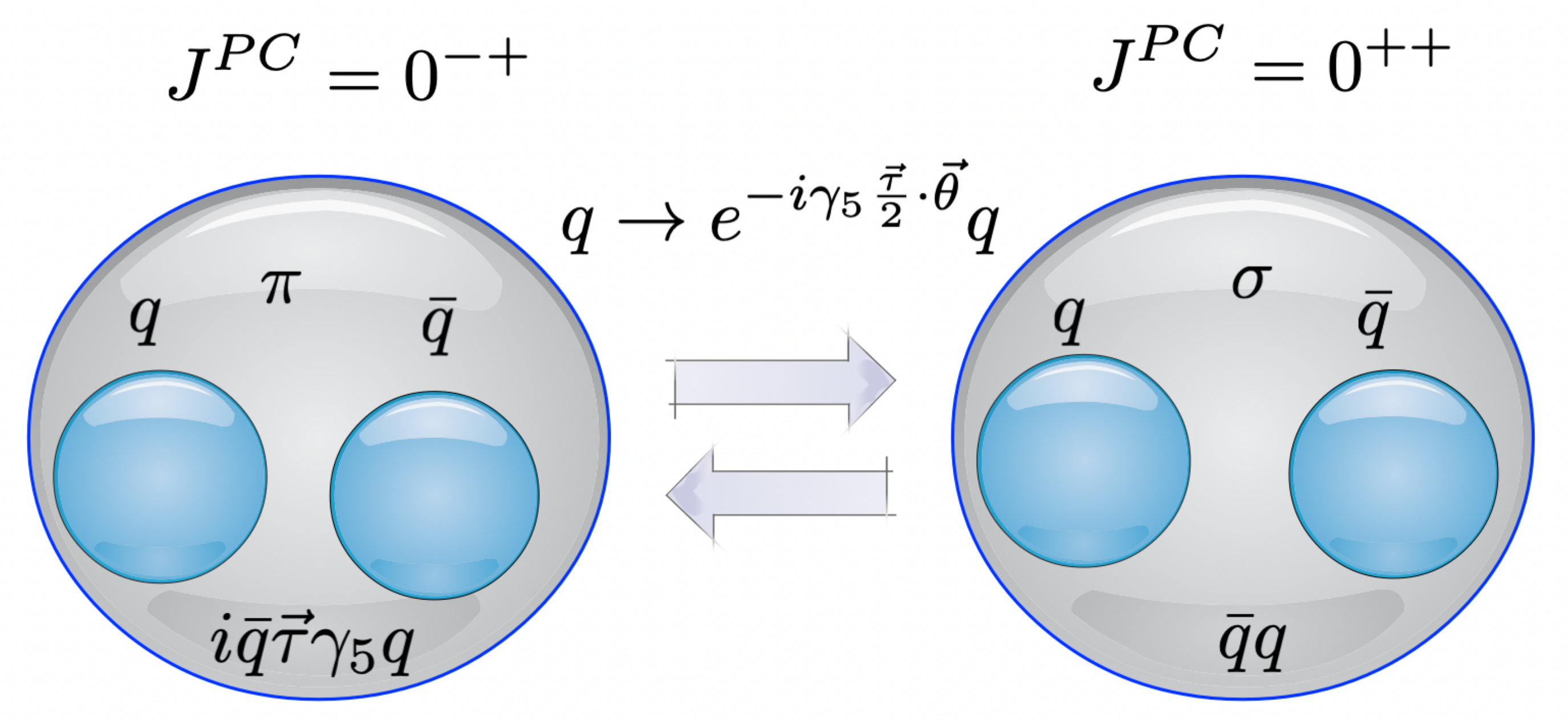}
       }
       \caption{\label{sca-spi}The S meson, e.g., $\sigma$ is viewed as the parity partner of the pion $\pi$. Note that the scalars in this article only refer to their quark-antiquark content.} 
\end{figure}

Recall that an S meson is a $0^{++}$ state. It can be considered as the chiral partner of the PS meson Fig.~\ref{sca-spi}. We work under the assumption that all states are purely quark-antiquark states. Then, for example, the states $\pi$ and $\sigma$ get transformed into each other through the following chiral transformation:
\bea
q \rightarrow {\rm e}^{-i \gamma_5 \frac{\bm{\tau}}{2} \cdot {\bm \theta}} q \,.
\eea
{\hspace{-0.3cm}} The explicit expression for the EFFs for S mesons with mass $M_M$ constituted from a quark $\fd$ and an antiquark $\bar\fdu$ is given by Eq.~(\ref{eqn:TotalMesonFF}) with
 \bea 
&& \hspace{-5mm} F^{\Ms,\fd}= 
P_T(Q^2)
E_{\Ms}^2 T^{\Ms}_{EE}(Q^2)  \,,\eea

\noindent
where
\begin{align}
\nn T^{\Ms}_{EE}(Q^2) & =- \frac{3}{4\pi^2}\bigg[ \int_0^1 d\alpha\, \overline{\mathcal{C}}_1(\omega_1) \\
 &
+ 
2\int_0^1 d\alpha\, d\beta \, \alpha \, \mathcal{A}^{\Ms}_{EE} \, \overline{\mathcal{C}}_2(\omega_2) \bigg]\,,
\end{align}

 \begin{table}[t!]
 \caption{ \label{Sparameters} 
 Ultraviolet regulator and the coupling constant for different combinations of quarks in S mesons. As before, $\hat{\alpha}_{\mathrm {IR}}=\hat{\alpha}_{\mathrm{IRL}}/Z_H$, where $\hat{\alpha}_{\mathrm {IRL}}=4.57$ is extracted from a best-fit to data as explained in Ref.~\cite{Raya:2017ggu}.  $\Lambda_{\rm IR} = 0.24$ GeV.} 
\begin{center}
\label{parametersSC}
\begin{tabular}{@{\extracolsep{0.0 cm}} || l | c | c | c ||}
\hline \hline
 \, quarks \, & $Z_{H}$ & \, $\Lambda_{\mathrm {UV}}\,[\GeV] \, $ & $\hat{\alpha}_{\mathrm {IR}}$ 
 \\
 \hline
 \rule{0ex}{2.5ex}
$\, \tu,\td,\ts$ & 1 & 0.905 & 4.57 \\ 
\rule{0ex}{2.5ex}
$\, \tc,\tu$ & 3.034 & 1.322 & 1.50 \\ 
\rule{0ex}{2.5ex}
$\, \tc,\ts$ & 3.034 & 2.222 & 1.50 \\ 
\rule{0ex}{2.5ex}
$\, \tc$     & 13.122 & 2.305 & 0.35 \\
\rule{0ex}{2.5ex}
$\, \tb,\tu$ & 18.473 & 10.670 & 0.25 \\
\rule{0ex}{2.5ex}
 $\, \tb,\ts$ & 29.537 & 11.064 & 0.15 \\
\rule{0ex}{2.5ex}
$\, \tb,\tc$   & 34.216 & 14.328 & 0.13 \\
\rule{0ex}{2.5ex}
$\, \tb$     & \, 127.013 \, & 26.873 & \, 0.036 \, \\
\hline \hline
\end{tabular}
\end{center}
\end{table}
\begin{table}[t!]
\caption{\label{par-scalar} Computed values of the S mesons masses and BSAs in the CI model, see~\cite{Gutierrez-Guerrero:2019uwa} for comparison, using the parameters listed in Tables~\ref{table-M} and \ref{parametersSC}.} 
\begin{tabular}{@{\extracolsep{0. cm}} || l | cc | c | c || }
\toprule
 \rule{0ex}{2.5ex}
 &  \, Mass [GeV] \, &  \,$E_{\Ms}$ \,  &  \,$m_{S}^{\rm exp}$ [GeV]  \,& \, error [\%]  \, \\ 
 \hline
 \rule{0ex}{2.5ex}
  $\tu\bar\td$ \, &  1.22  & 0.66 & --- & ---   \\
 \rule{0ex}{2.5ex}
 $\tu\bar\ts$ &  1.38  & 0.65 & --- & ---   \\
\rule{0ex}{2.5ex}
$\ts\bar{\ts}$ & 1.46  & 0.64 & --- & ---   \\
\rule{0ex}{2.5ex}
$\tc\bar{\tu}$ & 2.31  & 0.39 & 2.30  & 0.19   \\
\rule{0ex}{2.5ex}
$\tc\bar{\ts}$ & 2.42  & 0.42 & 2.32 & 3.54  \\
\rule{0ex}{2.5ex}
$\tu\bar{\tb}$ & 5.30  & 1.53 & --- & ---   \\
\rule{0ex}{2.5ex}
$\ts\bar{\tb}$ & 5.64  & 0.26 & --- & ---   \\
\rule{0ex}{2.5ex}
$\tc\bar{\tb}$ & 6.36  & 1.23 & 6.71 & 5.26   \\
\rule{0ex}{2.5ex}
$\tc\bar{\tc}$ & 3.33  & 0.16 & 3.42 & 2.73   \\
\rule{0ex}{2.5ex}
$\tb\bar{\tb}$ & 9.57  & 0.69 & 9.86 & 2.95   \\
\hline
\hline
\end{tabular}
\end{table}
%
\begin{figure*}[t!]
\begin{tabular}{@{\extracolsep{-2.3 cm}}c}
 \renewcommand{\arraystretch}{-1.6} %
 \hspace{-1cm}
 \includegraphics[scale=0.7]{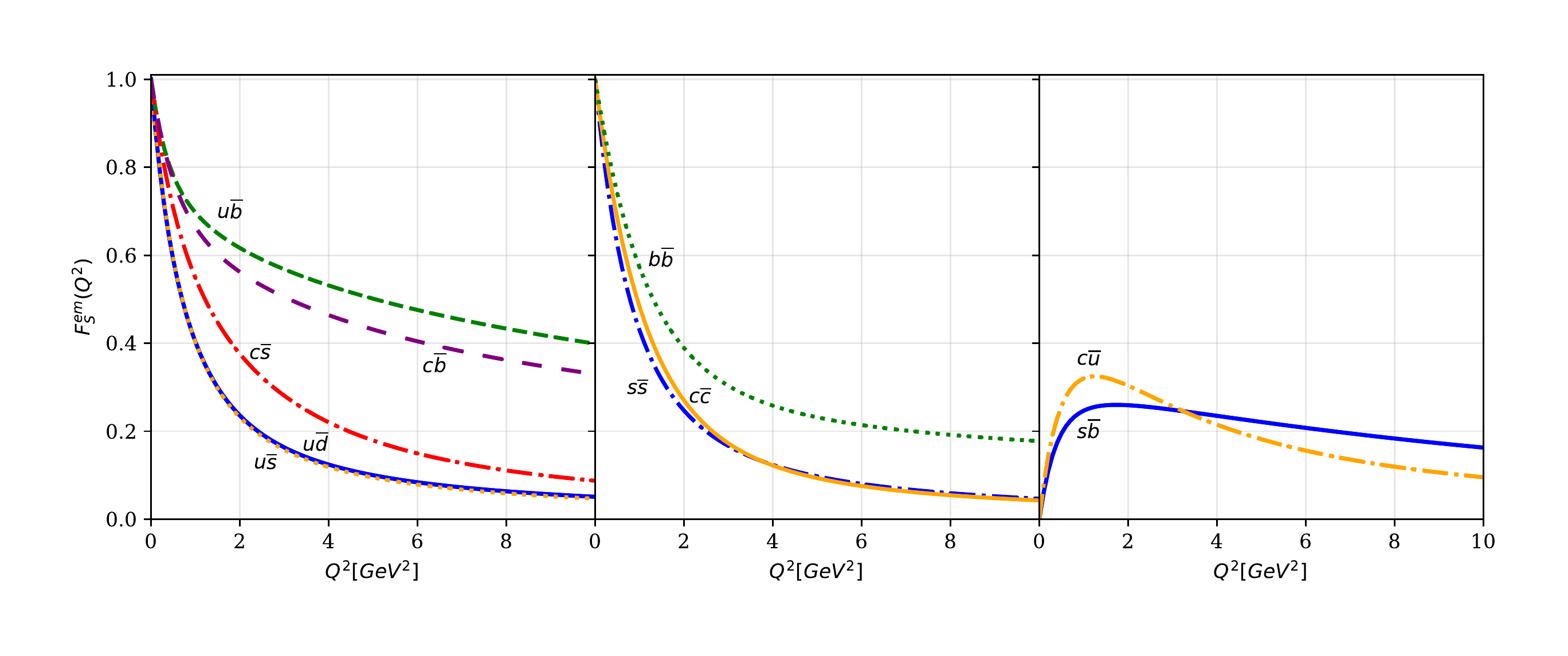}
\end{tabular}
\caption{EFFs for S mesons in the CI model. Left panel: electrically charges mesons composed of quarks of different flavors. Central panel: quarkonia including a hypothetical ground state {\em strangeonium} ($\ts\bar \ts$). Right panel: electrically neutral mesons composed of quarks of different flavors.
EFFs of electrically neutral but flavored mesons have been normalized to $F^S(0)=0$.}\label{plotS}
\end{figure*}
 with
 \begin{align}
     \mathcal{A}_{EE}^{\Ms} &= \alpha M_{\fd}-2(1-\alpha) M_{\fd} M_{\fu} \nonumber \\
     &+(\alpha-2)M_{\fu}^2+\alpha M_M^2 \, .
 \end{align}
\noindent
Note the close resemblance between $T_{EE}^{\Ms}$ and $T_{EE}^{\Meps}$. As expected, there are only sign differences between the two due to the presence, or absence, of the $\gamma^5$ matrix. 
In Table~\ref{Sparameters}, we present the parameters used for S mesons in order to compute the masses, amplitudes and charge radii.
We enlist the masses and BSAs of S mesons in Table~\ref{par-scalar} while the EFFs are depicted in Fig.~\ref{plotS}. On the right and central panels we present the results for neutral mesons while the left panel displays the EFFs of charged  mesons. We emphasize that for electrically neutral but flavored S mesons, we normalize the EFFs to zero at $Q^2=0$, while for flavorless mesons, the normalization is $F^S(0)=1$ to be consistent with the definition employed in Eq.~(\ref{eqn:TotalMesonFF}). We again perform a fit in the range $Q^2\in [0,8 M_M^2 ]$, where $M_M$ is the mass of the S meson. All the curves are faithfully reproduced by the following choice:
\begin{table*}[H]
\caption{\label{tableradiiPS} The charge radius for S mesons. All quantities are given in fm.}
\begin{center}
\begin{tabular}{@{\extracolsep{0.5 cm}}cccccccccccc}
\hline\hline
&$\tu\bar{\td}$  & $\tu\bar{\ts}$ & $\ts\bar{\ts}$ & $\tc\bar{\tu}$ & $\tc\bar{\ts}$  &$\tu\bar{\tb}$ & $\ts\bar{\tb}$ & $\tc\bar{\tb}$ & $\tc\bar{\tc}$ & $\tb\bar{\tb}$ \\
\rule{0ex}{2.5ex}
$r_{\rm S} $ & 0.55 & 0.54 & 0.50 & 0.46 & 0.42 &0.21 &0.41 &0.53 &0.43 & 0.43
\\ \hline\hline
\end{tabular}
\end{center}
\end{table*}
\begin{align}
F^{S}( Q^2) = \frac{e_{M} + a_S\, Q^2 +b_S\,Q^4}{1+c_S\, Q^2 + d_S\, Q^4}\,,
\label{EqfitS}
\end{align}
\begin{table}[t]
\caption{\label{fitParaS}Parameters for the fit in Eq.~(\ref{EqfitS}) for S mesons.}
\begin{tabular}{@{\extracolsep{0.0 cm}}||c|c|c|c|c||}
\hline
\hline
 &$a_S$ & $b_S$ & $c_S$ & $d_S$ \\ 
\hline
\rule{0ex}{2.5ex}
$\tu\bar{\td}$ \,\,& \,\,0.286\,\, &\,\, 0.003 \,\,&\,\, 1.543 \,\,& \,\,0.617\,\, \\
\rule{0ex}{2.5ex}
$\tu\bar{\ts}$ \,& 0.266 & 0.002 & 1.486 & 0.629  \\
\rule{0ex}{2.5ex}
$\ts\bar{\ts}$ \,& 0.217 & 0.001 & 1.271 & 0.542  \\
\rule{0ex}{2.5ex}
$\tc\bar{\tu}$ \,& 0.759 & $-$0.005 & 0.680 & 0.641 \\
\rule{0ex}{2.5ex}
$\tc\bar{\ts}$ \,& 0.004 & 0.001 & 0.783 & 0.047  \\
\rule{0ex}{2.5ex}
$\tu\bar{\tb}$ \,& 0.984 & 0.001 & 1.619 & 0.087  \\
\rule{0ex}{2.5ex}
$\ts\bar{\tb}$ \,& 0.210 & 0.001 & 0.175 & 0.115 \\
\rule{0ex}{2.5ex}
$\tc\bar{\tb}$ \,& 0.289 & 0.001 & 0.743 & 0.026  \\
\rule{0ex}{2.5ex}
$\tc\bar{\tc}$ \,& 0.217 & 0.001 & 0.860 & 0.673  \\
\rule{0ex}{2.5ex}
$\tb\bar{\tb}$ \,&0.269 & 0.000 & 1.607 & 0.020   \\ \hline \hline
\end{tabular}
\end{table}
\hspace{-0.3cm} where $e_M:=F^{S}(Q^2=0)$ is the electric charge of the meson and $a_S,b_S,c_S,d_S$ are the parameters of the fit. These values for S mesons are listed in Table \ref{fitParaS}. Based on these numbers, we can immediately infer the large $Q^2$ behavior of these EFFs. The coefficient $b_S \approx 0 $ for all S mesons under consideration. Therefore, the EFFs for S mesons fall as $1/Q^2$ for large $Q^2$. 

We present the numerical values of the charge radii for S mesons in \tab{tableradiiS}. 
We must reiterate that for the S mesons there are no reported measurements of their charge radii. Theoretical results are also scarce for any direct and meaningful comparison. It is worth mentioning again that the internal structure of scalar mesons is not well-established. Our results are based on  considering them as effective quark-antiquark states.


\begin{table*}[t!]
\begin{center}
\caption{\label{tableradiiS} The charge radii for S mesons. All quantities are reported in fm.}
 \renewcommand{\arraystretch}{1.6} %
\begin{tabular}{@{\extracolsep{0.0 cm}} ||c|c|c|c|c|c|c|c|c|c|c||}
\hline\hline
\rule{0ex}{2.5ex}
&$\tu\bar{\td}$  & $\tu\bar{\ts}$ & $\ts\bar{\ts}$ & $\tc\bar{\tu}$ & $\tc\bar{\ts}$  &$\tu\bar{\tb}$ & $\ts\bar{\tb}$ & $\tc\bar{\tb}$ & $\tc\bar{\tc}$ & $\tb\bar{\tb}$ \\
\hline
\rule{0ex}{2.5ex}
Our Result \,&\, $0.55$ \,&\, $0.54$ \,&\, $0.50$ \,&\, $0.47$ \,&\, $0.44$ \,&\,$0.42$ \,&\,$0.41$ \,&\, $0.40$ \,&\, $0.43$ \,&\, $0.39$\,\\
\hline\hline
\end{tabular}
\end{center}
\end{table*}

\begin{figure}[t]
\centerline{
 \includegraphics[scale=0.9,angle=0]{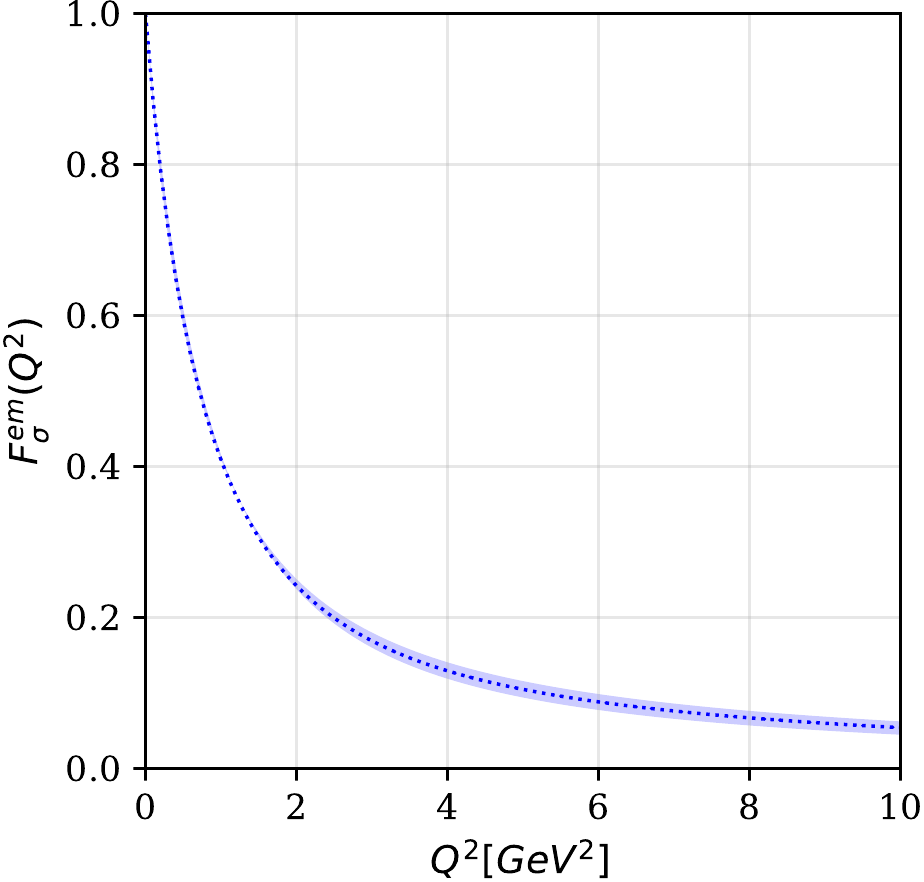}
 }
 \caption{ EFF for $\sigma$-meson. The central curve is obtained using the $\Lambda_{\rm UV}$ value from  \tab{parametersSC} while the band represents a $5\%$ variation in the charge radius.}
 \label{sigma}
\end{figure}

\begin{figure}[H]
    \centering
    \includegraphics[scale=0.47]{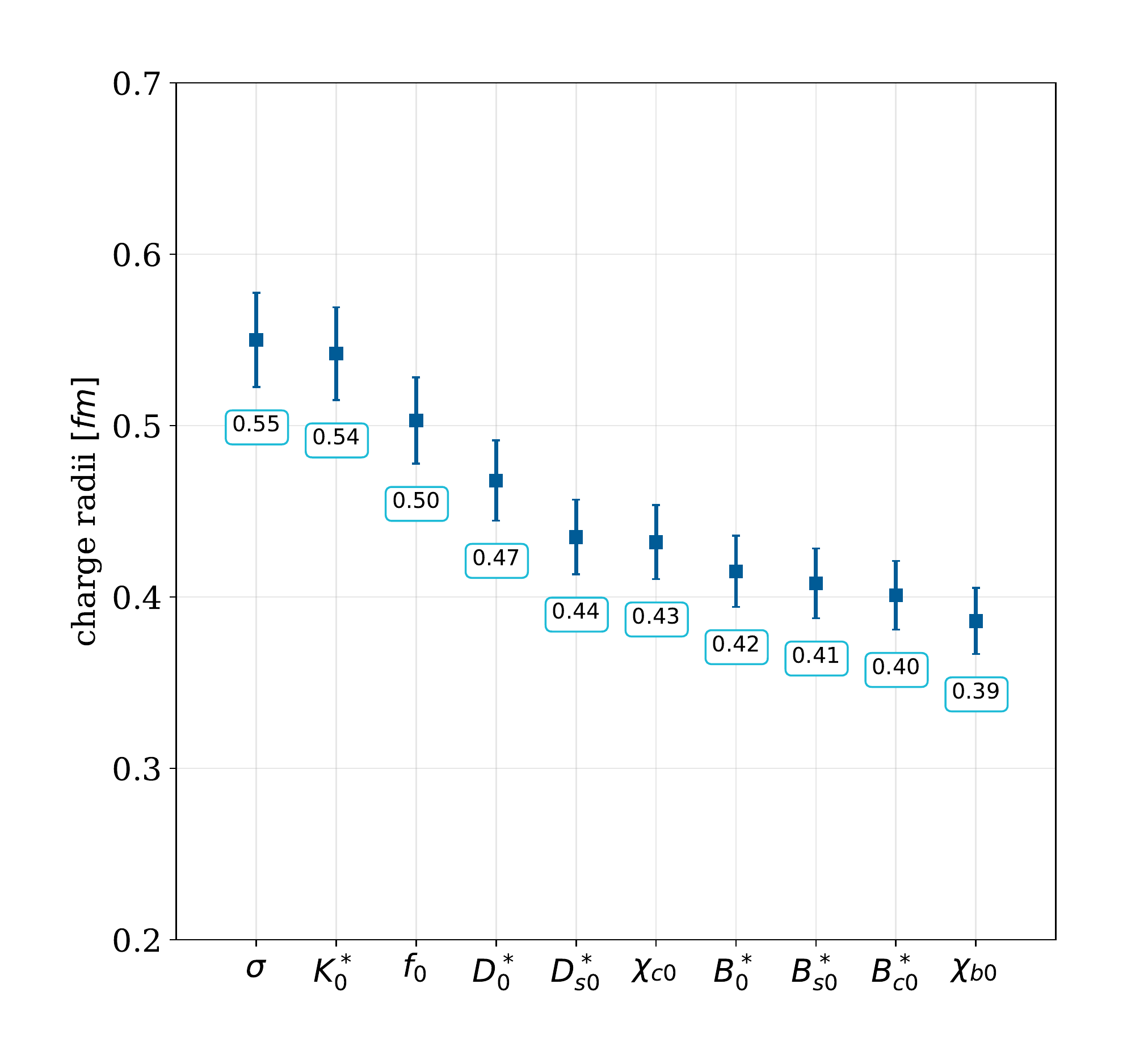}
    \caption{Charge radii of S mesons.}
    \label{fig:S_hierarchy}
\end{figure}

We would like to remind the reader that we again allow for a 5\% variation in the charge radii of S mesons. However, in 
Fig.~\ref{plotS}, we present the EFFs only for their central values for  visual clarity,  refraining from showing the corresponding band to avoid possible overlapping. However, in Fig.~\ref{sigma}, we depict a representative plot with a 5\% variation in the charge radius for the lightest scalar meson, $\sigma$, alone. Other mesons have similar bands.  
Finally, in Fig.~\ref{fig:S_hierarchy}, we plot the charge radii, extracted from the EFFs, as a function of the S meson mass. In general, the charge radii decrease when the S meson masses increase just as we observed for the PS mesons. 


\section{Conclusions}
\label{Conclusions}

In this work we present an exhaustive computation of EFFs employing the CI model for twenty ground state PS and S mesons. 
Note that the CI findings for light mesons and heavy quarkonia are already found in the literature as mentioned before~\cite{Gutierrez-Guerrero:2010waf,Bedolla:2015mpa,Bedolla:2016yxq,Raya:2017ggu}. We include these results for the sake of completeness and as a guide to pin down the best parameters in order to explore heavy-light systems. We thus report first results on the latter mesons within this model/formalism. We expect these new EFFs to be harder than the exact QCD predictions, especially for the PS mesons due to the necessary inclusion of the $F$-amplitude. We also anticipate the charge radii to be in the ballpark of a (20-25)\% error in light of the results where comparison 
with realistic studies and/or experiment has been possible. 


Furthermore, we analyze the sensitivity of the evolution of the EFFs by a change in appropriate parameters to allow for a 5\% variation in the charge radii of the corresponding mesons. The evolution band has been shown explicitly for $\pi, K, \sigma$ and $\eta_c$ alone to avoid over-crowding in other collective plots. However, it is worth mentioning that the corresponding bands in other EFFs are almost identical.
Interpolations have also been provided in Eqs.~(\ref{EqfitPS}, \ref{EqfitS}) and
Tables~\ref{fitParaPS} and~\ref{fitParaS} which allow for a convenient algebraic analysis of the behavior of the EFFs in the momentum range that we mentioned above and for any application the reader may deem useful.
We plan to  recalculate these EFFs for vector and axial vector mesons followed by the same computation within a more realistic algebraic model and in the long run for truncations more akin to full non-perturbative  QCD. It is also straightforward to generalise our analysis to study diquarks EFFs which are crucial in the subsequent computation of baryons EFFs such as the ones reported recently in~\cite{Raya:2021pyr}. All this is for future. \\
    
\begin{acknowledgements}
L.~X.~Guti\'errez-Guerrero wishes to thank the support from C\'atedras CONACyT program of Mexico. The work of R.~J.~Hern\'andez-Pinto is supported by CONACyT (Mexico) Project No. 320856 ({\em Paradigmas y Controversias de la Ciencia 2022}), {\em Ciencia de Frontera 2021-2042} and {\em Sistema Nacional de Investigadores} as well as by PROFAPI 2022 Grant No. PRO\_A1\_024 ({\em Universidad Aut\'onoma de Sinaloa}). The work of A.~Bashir is supported in part by the US Department of Energy (DOE) Contract No. DE-AC05-06OR23177, under which Jefferson Science Associates, LLC operates Jefferson Lab.
A.~Bashir also acknowledges {\em Coordinaci\'on de la Investigaci\'on Cient\'ifica} of the {\em Universidad Michoacana de San Nicol\'as de Hidalgo} grant 4.10 and the Fulbright-Garc\'ia Robles scholarship for his stay as a visiting scientist at the Thomas Jefferson National Accelerator Facility, Newport News, Virginia, USA. We thank Jozef Dudek and Christine Davies for helpful communication on lattice results on EFFs and charge radii. 

\end{acknowledgements}

\bibliography{ccc-a.bib}
\end{document}